\documentclass[twocolumn,showpacs,nofootinbib]{revtex4-1}

\usepackage[nointlimits]{amsmath}
\usepackage{hyperref}
\usepackage{amsfonts}
\usepackage{amssymb}
\usepackage{amsmath}
\usepackage{amsthm}
\usepackage{latexsym}
\usepackage{graphicx}
\usepackage{color}
\usepackage{layout}
\usepackage[english]{babel}
\def\LL{\mathcal{L}}
\def\x{\mathbf{x}}
\def\y{\mathbf{y}}
\def\z{\mathbf{z}}
\def\v{\mathbf{v}}
\def\V{\mathcal{V}}
\def\Y{\mathcal{Y}}

\allowdisplaybreaks

\begin{document}

\title{The $1\slash c$ expansion of nonminimally coupled curvature-matter gravity model and constraints from
planetary precession}

\author{Riccardo March}
\email{r.march@iac.cnr.it}
\affiliation{Istituto per le Applicazioni del Calcolo, CNR, Via dei Taurini 19, 00185 Roma, Italy}
\affiliation{INFN - Laboratori Nazionali di Frascati (LNF), Via E. Fermi 40, Frascati 00044 Roma, Italy}

\author{Jorge P\'aramos}
\email{jorge.paramos@fc.up.pt}

\author{Orfeu Bertolami}
\email{orfeu.bertolami@fc.up.pt}
\affiliation{Departamento de F\'isica e Astronomia and Centro de F\'isica do Porto,\\Faculdade de Ci\^encias da Universidade do Porto, Rua do Campo Alegre 687, 4169-007 , Porto, Portugal}

\author{Simone Dell'Agnello}
\email{simone.dellagnello@lnf.infn.it}
\affiliation{INFN - Laboratori Nazionali di Frascati (LNF), Via E. Fermi 40, Frascati 00044 Roma, Italy}

\date{\today}
\begin{abstract}
The effects of a nonminimally coupled curvature-matter model of gravity on a perturbed Minkowski metric are presented. The action functional of the model involves two functions $f^1(R)$ and $f^2(R)$ of the Ricci scalar curvature $R$.
This work expands upon previous results, extending the framework developed there to compute corrections up to order $O\left(1\slash c^4\right)$ of the 00 component of the metric tensor.
It is shown that additional contributions arise due to both the non-linear form $f^1(R)$ and the nonminimal coupling $f^2(R)$, including exponential contributions that cannot be expressed as an expansion in powers of $1/r$. Some possible experimental implications are assessed with application to perihelion precession.

\end{abstract}

\pacs{04.20.Fy, 04.80.Cc, 04.25.Nx}

\maketitle

\section{Introduction}

Dark matter and dark energy are key contemporary concepts used to account, for instance, for the astrophysical problem of the flattening of galactic rotation curves and the cosmological issue of the accelerated expansion of the universe, respectively. Dark energy accounts for 69$\%$ of the energy budget of the universe \cite{Planck}; among several other proposals, it has been the object of several so-called "quintessence" models \cite{quintessence}, which posit the existence of scalar fields with negative pressure, as an alternative to a suitably adjusted Cosmological Constant, which presents the eponymous problem of reconciling the large order of magnitude difference between its observed and predicted values \cite{Weinberg}. Dark matter searches focus on the characterization of additional matter species arising from extensions to the Standard Model of particles, collectively dubbed as weak-interacting massive particles (WIMPS) such as, for instance, neutralinos or axions \cite{DM}. As an alternative, some proposals assume that both dark components may be described in a unified fashion \cite{scalarfield,Chaplygin}.

Other models assume that, instead of additional matter species, the fundamental laws of General Relativity (GR) may be incomplete, prompting {\it e.g.} for corrections and alternatives to the Einstein-Hilbert action. Among such theories, those involving a nonlinear corrections to the geometric part of the action via the scalar curvature, aptly  called $f(R)$ theories, have gained much attention (see Ref. \cite{felice} for a thorough discussion). These can be extended also to include a nonminimal coupling (NMC) between the scalar curvature and the matter Lagrangian density, leading to an even richer phenomenology and implying that the energy-momentum tensor may not be (covariantly) conserved \cite{BBHL} (see also Ref. \cite{Lobo} for a more general model).

NMC models have yielded several interesting results, including the impact on stellar observables \cite{stelobserv}, energy conditions \cite{energcondit}, equivalence with multi-scalar-tensor theories \cite{multiscalar}, possibility to account for galactic \cite{drkmattgal} and cluster \cite{drkmattclus} dark matter, cosmological perturbations \cite{cosmpertur}, a mechanism for mimicking a Cosmological Constant at astrophysical scales \cite{mimlambda}, post-inflationary reheating \cite{reheating}, dark energy \cite{curraccel,Friedmann,Ribeiro}, dynamical impact of the choice of the Lagrangian density of matter \cite{dynimpac1,dynimpac2}, gravitational collapse \cite{gravcollapse} and black hole solutions \cite{BertolamiCadoni}, its Newtonian limit \cite{newtlimit}, the existence of closed timelike curves \cite{closedtimecurve} and the modified Layzer-Irvine equation \cite{LayzerIrvine} (see Ref. \cite{review} for a review and Refs. \cite{puetzobukiorio} for other NMC gravity theories and their potential applications).

Recently, the impact of NMC gravity on the spacetime metric surrounding a spherical central body was considered in Ref. \cite{BMP}, where the additional degree of freedom arising from a non-trivial $f(R)$ function is light, thus yielding a long-range additional force which requires considering the background cosmological setting; following the procedure set out in Ref. \cite{CSE} for $f(R)$ gravity, the Parameterized Post-Newtonian (PPN) parameter $\gamma $ was computed, provided that a set of requirements for $f(R)$ and the NMC function are obeyed.
Then the compatibility has been assessed between a NMC model which accounts for the observed accelerated expansion of
the Universe and Solar System experiments.

Conversely, the case where the former is short-ranged enables one to neglect the background cosmological setting and derive the ensuing corrections to the gravitational potential \cite{CPM}, which are shown to be of the Yukawa-type --- as previously reported in Ref. \cite{NJ} for $f(R)$ gravity. In particular, it is found that the range of this Yukawa potential is given solely by $f(R)$, with the NMC affecting only its strength: this is a natural result, since the effect of the latter vanishes in vacuum, but affects the gravitational source.

The purpose of this work is thus to further examine those findings, extending the formalism used in Ref. \cite{CPM} to include terms up to order $O\left(1/c^4\right)$ in the 00 component of the metric tensor. The nonlinear correction
to the geometry part of the action is represented by a function $f^1(R)$, and the NMC is represented by a function $f^2(R)$ which multiplies the matter Lagrangian density.
Both functions are assumed analytic at $R=0$ and the coefficients of the Taylor expansions around $R=0$
are considered as the parameters of the model.

This work is organized as follows: In section II, the NMC model is presented and in section III its nonrelativistic limit is derived. Section IV computes the post-Newtonian and Yukawa corrections to the metric tensor by considering
matter as a perfect fluid (without assumptions of symmetry). In Section V the metric around a static,
spherically symmetric body is computed. Section VI addresses
the ensuing Solar System constraints, namely through perturbations to perihelion precession.
Recent observations of Mercury, including data from the Messenger spacecraft, are used to constrain the parameters
of the model. Finally, conclusions are drawn in Section VII.


\section{Nonminimally Coupled Gravity}

The action functional of NMC gravity is of the form
\cite{BBHL}
\begin{equation}
S = \int \left[\frac{1}{2}f^1(R) + [1 + f^2(R)] \LL \right]\sqrt{-g}   d^4x,
\end{equation}
where $f^i(R)$ (with $i=1,2$) are functions of the Ricci scalar curvature $R$, $\LL$ is the Lagrangian
density of matter, and $g$ is the metric determinant.

The Einstein-Hilbert action is recovered by choosing:
\begin{equation}
f^1(R) = 2\kappa (R - 2\Lambda), \qquad f^2(R) = 0,
\end{equation}
where $\kappa \equiv c^4/16\pi G$, $G$ is Newton's gravitational constant and $\Lambda$ the cosmological constant.

The variation of the action functional with respect to the metric $g_{\mu\nu}$ yields the field equations
\begin{eqnarray}\label{field-eqs}
&& \left(f^1_R + 2f^2_R \LL \right) R_{\mu\nu} - \frac{1}{2} f^1 g_{\mu\nu} = \\ && \nonumber
\left(\nabla_\mu \nabla_\nu -g_{\mu\nu} \square \right) \left(f^1_R + 2f^2_R \LL \right)
+ \left(1 + f^2 \right) T_{\mu\nu},
\end{eqnarray}
where $f^i_R \equiv df^i\slash dR$. The trace of the field equations is given by
\begin{eqnarray}
\nonumber && \left( f^1_R + 2f^2_R \LL \right) R + 3\square ( f^1_R +
2f^2_R \LL ) - 2f^1 = \\  && \left( 1 + f^2 \right) T,\label{trace}
\end{eqnarray}
where $T$ is the trace of the energy-momentum tensor $T_{\mu\nu}$.

A rather striking feature of NMC gravity is that the energy-momentum tensor of matter is not covariantly conserved: indeed, applying the Bianchi identities to Eq. (\ref{field-eqs}), one finds that
\begin{equation}
\nabla_\mu T^{\mu\nu} = \frac{f^2_R }{ 1 + f_2} ( g^{\mu\nu} \LL - T^{\mu\nu} ) \nabla_\mu R,
\end{equation}
a result that, as discussed thoroughly in Refs. \cite{multiscalar,Sotiriou1}, cannot be ``gauged away'' by a convenient conformal transformation, but is instead a distinctive feature of the model under scrutiny.
\subsection{Assumptions on the metric}

We assume that the metric can be written as a small perturbation around flat spacetime,
\begin{equation}
g_{\mu\nu} = \eta_{\mu\nu} + h_{\mu\nu}, \qquad \mbox{with } \left| h_{\mu\nu} \right| \ll 1,
\end{equation}
where $\eta_{\mu\nu}$ is the Minkowski metric with signature $(-,+,+,+)$. In the following, Greek letters
denote space-time indices ranging from 0 to 3, whereas Latin letters denote spatial indices ranging from 1 to 3.

In analogy with the post-Newtonian approximation of General Relativity, we expand the metric tensor
in powers of $1\slash c$:
\begin{eqnarray}
g_{00} &=& -1 + h_{00}^{(2)} + h_{00}^{(4)} + O\left(\frac{1}{c^6}\right),\\ \nonumber
g_{0i} &=& h_{0i}^{(3)} + O\left(\frac{1}{c^5}\right),\\ \nonumber
g_{ij} &=& \delta_{ij} + h_{ij}^{(2)} + O\left(\frac{1}{c^4}\right),
\end{eqnarray}
where
\begin{equation}
h_{\mu\nu}^{(n)} = O\left(\frac{1}{c^n}\right), \qquad\mbox{for } n=2,3,4.
\end{equation}
We impose the following gauge conditions \cite{Wi},
\begin{eqnarray}
h_{i0,i}^{(3)} &=& \frac{1}{2c}  h_{ii,0}^{(2)} + O\left(\frac{1}{c^5}\right), \label{gauge-1}\\ \nonumber
h_{ij,j}^{(2)} &=& \frac{1}{2}  h_{jj,i}^{(2)} - \frac{1}{2}  h_{00,i}^{(2)} + O\left(\frac{1}{c^4}\right),
\end{eqnarray}
so that the Ricci tensor $R_{\mu\nu}$ is expanded as
\begin{eqnarray}\label{R-00}
R_{00} &=& - \frac{1}{2}  \nabla^2 h_{00}^{(2)} - \frac{1}{2}  \nabla^2 h_{00}^{(4)}
- \\ \nonumber && - \frac{1}{2}  \left| \nabla h_{00}^{(2)} \right|^2 + \frac{1}{2}  h_{ij}^{(2)} h_{00,ij}^{(2)}
+ O\left(\frac{1}{c^6}\right), \\
R_{0i} &=& - \frac{1}{2}  \nabla^2 h_{0i}^{(3)} - \frac{1}{4c}  h_{00,i0}^{(2)}
+ O\left(\frac{1}{c^5}\right), \label{R-0i}\\ \label{R-ij}
R_{ij} &=& - \frac{1}{2}  \nabla^2 h_{ij}^{(2)} + O\left(\frac{1}{c^4}\right),
\end{eqnarray}
where $\nabla^2$ denotes the usual Laplacian operator in three-dimensional Euclidean space.

We also expand the Ricci scalar as follows:
\begin{equation}\label{R-expansion}
R = R^{(2)} + R^{(4)} + O\left(\frac{1}{c^6}\right),
\end{equation}
where $R^{(n)} = O\left(1\slash c^n\right)$, for $n=2,4$.

\subsection{Energy-momentum tensor}

As in the PPN framework, the components of the energy-momentum tensor, $T_{\mu\nu}$, to the relevant order, are \cite{Wi}
\begin{eqnarray}
\label{T-ij}
T_{00} &=& \rho c^2 \left( 1 + \frac{v^2}{c^2} + \frac{\Pi}{c^2} - h_{00}^{(2)} \right)
+ O\left(\frac{1}{c^2}\right), \label{T-00} \\
T_{0i} &=& -\rho c v_i + O\left(\frac{1}{c}\right), \label{T-0i}\\
T_{ij} &=& \rho v_i v_j + p\delta_{ij} + O\left(\frac{1}{c^2}\right), \end{eqnarray}
where matter is considered as a perfect fluid with matter density $\rho$, velocity field $v_i$, pressure $p$,
and specific energy density $\Pi$ (ratio of energy density to rest-mass density). The trace of the energy-momentum
tensor is given by
\begin{equation}
T = -\rho c^2 \left( 1 + \frac{\Pi}{c^2} \right) +3p + O\left(\frac{1}{c^2}\right).
\end{equation}
If $\Omega$ denotes the portion of three-dimensional space occupied by a body with mass density $\rho$,
and $\rho=0$ outside of the body, in order for the field Eqs. (\ref{field-eqs}) to be well defined, we require
that both the function $\rho=\rho(t,x)$ and its spatial derivatives are continuous across the surface of the body:
\begin{equation}\label{boundary-cond}
\rho(t,x)=0, \quad \nabla\rho(t,x)=0, \qquad x\in\partial\Omega,
\end{equation}
where the operator $\nabla$ denotes the three-dimensional gradient.
\\

In what follows, we use $\LL = -\rho c^2$ for the Lagrangian density of matter (see Ref. \cite{dynimpac1} for a discussion).

\subsection{Assumptions on $f^1(R)$ and $f^2(R)$}

We assume the functions $f^1(R)$ and $f^2(R)$ to be analytic at $R=0$. Hence, the function $f^1$ admits
the following Taylor expansion around $R=0$,
\begin{equation}\label{expansionf1}
f^1(R) = 2\kappa \sum_{i=1}^\infty a_i R^i, \qquad a_1=1,
\end{equation}
where the condition $a_1=1$ allows for recovering GR when the function $f^1$ is linear and $f^2=0$.

Analogously, the function $f^2$ admits the following Taylor expansion,
\begin{equation}\label{expansionf2}
f^2(R) = \sum_{j=1}^\infty q_j R^j.
\end{equation}
The $1\slash c$ expansion of the metric, which is the subject of the present paper,
will show how the coefficients $a_i,q_j$ affect the weak-field limit of NMC gravity,
in such a way that some of these coefficients can be constrained by means of experiments in gravitational physics.

\section{Nonrelativistic limit}\label{sec:nonrel-limit}

In this section we compute the quantity $h_{00}^{(2)}$, which yields the nonrelativistic limit
of NMC gravity. First, we compute the trace of the field Eqs. (\ref{trace}) at order $O\left(1\slash c^2\right)$, obtaining
\begin{equation}\label{trace-2}
\nabla^2 R^{(2)} - \frac{R^{(2)} }{6a_2}= -\frac{4\pi G}{3c^2 a_2}
\left( \rho - 6q_1\nabla^2\rho \right).
\end{equation}
In the following we assume that $a_2>0$ and set $m^2 = 1\slash(6a_2)$.

The above admits a Yukawa-type solution,
\begin{eqnarray}
&& R^{(2)} = \frac{G}{3c^2a_2} \times \\ \nonumber && \int d^3y \frac{e^{-m|\x-\y|}}{|\x-\y|} \left[ \rho(t,\y) -6q_1\nabla^2\rho(t,\y) \right].
\end{eqnarray}
We now introduce the Green function
\begin{equation}
G(\x-\y) = -\frac{1}{4\pi} \frac{e^{-m|\x-\y|}}{|\x-\y|},
\end{equation}
which satisfies the following equation in the sense of a distribution,
\begin{equation}
( \nabla^2 - m^2 ) G(\x-\y) = \delta(\x-\y),
\end{equation}
where $\delta({\bf x} - {\bf y})$ is the Dirac distribution.

Hence, if the mass density $\rho$ is zero outside of a body which occupies a region $\Omega$ of three-dimensional space, using Green's identity and the boundary conditions Eq. (\ref{boundary-cond}), we have
\begin{eqnarray}
&& \int \nabla^2\rho(t,\y)\frac{e^{-m|\x-\y|}}{|\x-\y|} d^3y = \\ \nonumber &&
 - 4\pi\rho + m^2 \int \rho(t,\y)\frac{e^{-m|\x-\y|}}{|\x-\y|} d^3y.
\end{eqnarray}
Collecting the above results we find for the Ricci scalar $R$ at order $O\left(1\slash c^2\right)$:
\begin{eqnarray}\label{R2-solution}
&& R^{(2)} =\frac{8\pi G}{c^2} \frac{q_1}{a_2}\rho + \\ \nonumber &&  \frac{G}{3c^2a_2}\left( 1 - \frac{q_1}{a_2} \right)\int\rho(t,\y)\frac{e^{-m|\x-\y|}}{|\x-\y|} d^3y.
\end{eqnarray}
Note that, if $a_2<0$, then the solution for $R^{(2)}$ would be oscillatory, which would lead to an unphysical behaviour at asymptotically large distances.

The $0-0$ component of the field Eqs. (\ref{field-eqs}), written at order $O\left(1\slash c^2\right)$, is
\begin{equation}\label{field-eqs-00-2}
\nabla^2 \left( h_{00}^{(2)} + 4 a_2 R^{(2)} - \frac{2q_1}{\kappa} \rho c^2\right) =
R^{(2)} - \frac{1}{\kappa}\rho c^2,
\end{equation}
where the $O\left(1\slash c^2\right)$ contributions to $R_{00}$ and $T_{00}$ have been taken
into account using Eqs. (\ref{R-00}) and (\ref{T-00}), respectively.

Combining Eq. (\ref{field-eqs-00-2}) with the trace Eq. (\ref{trace-2}) yields the modified Poisson equation
\begin{equation}
\nabla^2 \left( h_{00}^{(2)} -2a_2 R^{(2)} + \frac{16\pi G}{c^2}  q_1\rho \right) =
-\frac{8\pi G}{c^2} \rho,
\end{equation}
which admits the solution
\begin{equation}\label{h00-2}
h_{00}^{(2)} = 2 \left( \frac{U}{c^2} + a_2 R^{(2)} - \frac{8\pi G}{c^2}  q_1\rho \right),
\end{equation}
where $U$ is the usual Newtonian potential
\begin{equation}
U = G \int \frac{\rho(t,\y)}{|\x-\y|} d^3y.
\end{equation}
In the particular case of a body with a static and spherically symmetric distribution of mass, the solution Eq. (\ref{h00-2}) coincides, outside of the body, with the metric found in Ref. \cite{CPM}; in the case of pure $f(R)$ gravity, {\it i.e.} $q_1=0$, it reduces to the solution for $h_{00}^{(2)}$ found in Ref. \cite{Cl}.

Eventually, the solution for $h_{00}^{(2)}$ shows that the nonrelativistic limit of NMC gravity,
outside of a massive body, is constituted by the sum of the Newtonian potential plus a Yukawa potential
proportional to $R^{(2)}$. The characteristic length of the Yukawa potential is given by $\lambda \equiv 1\slash m$,
as in $f(R)$ gravity, whereas the strength of such a potential depends on both $a_2$ and the NMC parameter $q_1$.

The gravitational effects of this Yukawa potential and consequent experimental constraints on the parameters $a_2$ and $q_1$ have been discussed in detail in Ref. \cite{CPM}.

\section{Post-Newtonian + Yukawa approximation of NMC gravity}\label{sec:PPNY}

In this section we compute a {\it parametrized post-Newton plus Yukawa} (PPNY) approximation of NMC gravity
(see also \cite{NJ} for $f(R)$ gravity): this reflects the impossibility of expanding a
Yukawa perturbation $\sim (1/r)\exp(-r/\lambda)$ in powers of $1/r$, so that both contributions must be considered.
More precisely, in the following subsections we compute the metric contributions $h_{ij}^{(2)}, h_{0i}^{(3)}$ and $h_{00}^{(4)}$, by solving the field equations of NMC gravity.
\subsection{Solution for $h_{ij}$ at second order}

The $i-j$ components of the field Eqs. (\ref{field-eqs}), written at order $O\left(1\slash c^2\right)$, are
\begin{eqnarray}
\nonumber && \nabla^2 \left( \frac{1}{2} h_{ij}^{(2)} - 2a_2\delta_{ij}R^{(2)} +
\frac{16\pi G}{c^2} q_1\rho \delta_{ij} \right)
+  \\ && \frac{1}{2} \delta_{ij}R^{(2)} + 2a_2R^{(2)}_{ ,ij} = \frac{c^2}{\kappa} q_1 \rho_{,ij} , \label{field-eqs-ij-2}
\end{eqnarray}
where the $O\left(1\slash c^2\right)$ contributions to $R_{ij}$ and $T_{ij}\slash(2\kappa)$ have been taken
into account using Eqs. (\ref{R-ij}) and (\ref{T-ij}), respectively.

In order to rewrite Eq. (\ref{field-eqs-ij-2}) in the form of a Poisson equation, we observe that,
using Eqs. (\ref{R-00}) and (\ref{R-ij}) at order $O\left(1\slash c^2\right)$, we have
\begin{equation}\label{R2rewritten}
R^{(2)} = \frac{1}{2} \left( \nabla^2 h_{00}^{(2)} - \nabla^2 h_{ii}^{(2)} \right).
\end{equation}
Using this result and the $0-0$ component of the field Eqs. (\ref{field-eqs-00-2}), the trace Eq. (\ref{trace-2}) can be rewritten as
\begin{equation}\label{trace-rewritten}
\nabla^2 \left( h_{ii}^{(2)} + 5h_{00}^{(2)} \right) = -\frac{64\pi G}{c^2} \rho.
\end{equation}
Moreover, using the Poisson equation for the Newtonian potential, $\nabla^2U=-4\pi G\rho$, we have
\begin{equation}\label{rho-hessian}
\rho_{,ij} = - \frac{1}{4\pi G} \nabla^2 U_{,ij},
\end{equation}
while the solution (\ref{h00-2}) for $h_{00}^{(2)}$ and Eqs. (\ref{R2rewritten})-(\ref{rho-hessian}) enable to write
\begin{equation}\label{R2hessian}
R^{(2)}_{ ,ij} = \nabla^2 \left( 6a_2R^{(2)}_{ ,ij} - \frac{2}{c^2}U_{,ij}
- \frac{48\pi G}{c^2}q_1\rho_{,ij}  \right).
\end{equation}
Now, substituting Eqs. (\ref{rho-hessian}) and (\ref{R2hessian}) into the $i-j$ components of the field Eqs. (\ref{field-eqs-ij-2}), and using again Eq. (\ref{R2rewritten}) of $R^{(2)}$,
we obtain the following,
\begin{eqnarray}
&&\nabla^2 \bigg[ \frac{1}{2} h_{ij}^{(2)} + a_2\delta_{ij}R^{(2)} + 12a_2^2R^{(2)}_{ ,ij} - \\ \nonumber &&
\frac{4}{c^2}\left(a_2-q_1\right)U_{,ij} - \frac{8\pi G}{c^2} q_1 \left( \rho \delta_{ij} -
12  a_2\rho_{,ij} \right) \bigg]  \\ \nonumber && = - \frac{4\pi G}{c^2} \rho \delta_{ij}.
\end{eqnarray}
This is a system of decoupled Poisson equations with solution
\begin{eqnarray}\label{hij-2}
&& h_{ij}^{(2)} = 2 \bigg[ \frac{U}{c^2} \delta_{ij} - a_2\delta_{ij}R^{(2)} -
12a_2^2R^{(2)}_{ ,ij} + \\ \nonumber && \frac{4}{c^2}\left(a_2-q_1\right)U_{,ij} +
\frac{8\pi G}{c^2} q_1 \left( \rho \delta_{ij} + 12 a_2 \rho_{,ij} \right) \bigg].
\end{eqnarray}
In the case of pure $f(R)$ gravity, {\it i.e.} if $q_1=0$, the above reduces to the solution for $h_{ij}^{(2)}$ found in Ref. \cite{Cl}.

Notice that the obtained solution is not diagonal, and hence it is not in the standard post-Newtonian gauge. In a subsequent section, it will be written as a diagonal spatial metric by means of a suitable gauge
transformation.

\subsection{Solution for $h_{0i}$ at third order}

The $0-i$ components of the field Eqs. (\ref{field-eqs}), written at order $O\left(1\slash c^3\right)$, are
\begin{equation}\label{field-eqs-0i-3}
\nabla^2 h_{0i}^{(3)} + \frac{1}{2c} h^{(2)}_{00,0i} + \frac{4a_2}{c} R^{(2)}_{ ,0i} -
\frac{2c}{\kappa} q_1\rho_{,0i} = \frac{c}{\kappa} \rho v_i,
\end{equation}
where the $O\left(1\slash c^3\right)$ contributions to $R_{0i}$ and $T_{0i}$ have been taken
into account using Eqs. (\ref{R-0i}) and (\ref{T-0i}), respectively.

In order to solve Eqs. (\ref{field-eqs-0i-3}) we use the following set of PPN potentials \cite{Wi},
\begin{eqnarray}
V_i &=& G \int \frac{\rho(t,\y)v_i(t,\y)}{|\x-\y|} d^3y, \\ \nonumber
W_i &=& G \int \frac{\rho(t,\y)[\v(t,\y)\cdot(\x-\y)](x-y)_i}{|\x-\y|^3} d^3 y.
\end{eqnarray}
Using the continuity equation
\begin{equation}\label{contin-eq}
\frac{\partial\rho}{\partial t} + \nabla \cdot (\rho\v) = 0,
\end{equation}
one can show that ({\it cf.} Ref. \cite{Wi})
\begin{equation}\label{WiVi}
\nabla^2 \left( W_i - V_i \right) = 2 U_{,0i}.
\end{equation}
Then, arguing as in the previous subsection, we have
\begin{eqnarray}
R^{(2)}_{ ,0i} &=& \nabla^2 \left( 6a_2R^{(2)}_{ ,0i} - \frac{2}{c^2}U_{,0i}
- \frac{48\pi G}{c^2}q_1\rho_{,0i}  \right), \nonumber \\ \label{R20i-rho0i}
\rho_{,0i} &=& - \frac{1}{4\pi G} \nabla^2 U_{,0i}.
\end{eqnarray}
Inserting Eqs. (\ref{WiVi}) and (\ref{R20i-rho0i}) into the $0-i$ components of the field Eqs. (\ref{field-eqs-0i-3}) and using the solution (\ref{h00-2}) for $h_{00}^{(2)}$, we obtain
\begin{eqnarray}
\nabla^2 && \bigg[ h_{0i}^{(3)} + 30\frac{a_2^2}{c} R^{(2)}_{ ,0i} -
\frac{10}{c^3}\left(a_2-q_1\right)U_{,0i}  \nonumber \\ \nonumber && -\frac{1}{2c^3} V_i + \frac{1}{2c^3} W_i
-\frac{240\pi G}{c^3} a_2q_1\rho_{,0i} \bigg] \\
& & = \frac{16\pi G}{c^3} \rho v_i.
\end{eqnarray}
This is a system of decoupled Poisson equations with solution
\begin{eqnarray}
h_{0i}^{(3)} &=& -\frac{7}{2c^3} V_i - \frac{1}{2c^3} W_i + \frac{10}{c^3}\left(a_2-q_1\right)U_{,0i} \nonumber \\  && -
30\frac{a_2^2}{c} R^{(2)}_{ ,0i} + \frac{240\pi G}{c^3} a_2q_1\rho_{,0i}. \label{h0i-3}
\end{eqnarray}
Again, in the case of pure $f(R)$ gravity the above reduces to the solution for $h_{0i}^{(3)}$ found in Ref. \cite{Cl}.

\subsection{Solution for $h_{00}$ at fourth order}\label{subsec:h004}

The solution of the $0-0$ component of the field Eqs. (\ref{field-eqs})
at order $O\left(1\slash c^4\right)$ is more involved and its computation is deferred to Appendix A, leading to the lengthy expression shown below,
\begin{widetext}
\begin{eqnarray}\label{h004}
h_{00}^{(4)} &=& -\frac{2}{c^4} U^2 - 2a_2^2R^2 -4\frac{a_2}{c^2} UR +
\frac{32\pi G}{c^4} q_1\rho U + \frac{32\pi G}{c^2}a_2q_1\rho R \\
&-& \frac{128\pi^2G^2}{c^4} q_1^2\rho^2 - 36\frac{a_2^2}{c^2} R_{,00} + \frac{12}{c^4}\left(a_2-q_1\right)U_{,00}
+\frac{288\pi G}{c^4} a_2q_1\rho_{,00} \nonumber\\
&+& \frac{8}{c^4}\left(a_2-q_1\right)|\nabla U|^2 - 24a_2^3|\nabla R|^2 -
\frac{1536\pi^2G^2}{c^4} a_2q_1^2|\nabla\rho|^2 -
8\frac{a_2}{c^2}\left(2a_2+q_1\right)\nabla U \cdot \nabla R \nonumber\\
&+& \frac{64\pi G}{c^4} q_1\left(2a_2+q_1\right) \nabla\rho \cdot \nabla U +
\frac{384\pi G}{c^2} a_2^2q_1 \nabla\rho \cdot \nabla R
-\frac{a_2}{3\pi} \V(R^2) + \frac{4G}{c^4} \V(\rho U) \nonumber\\
&-& \frac{2G}{c^2}\left(\frac{8}{3} a_2-5q_1\right)\V(\rho R)
+ \frac{64\pi G^2}{c^4} q_1\V(\rho^2) - \frac{8G}{c^4}\left(a_2-q_1\right)\V(\nabla\rho \cdot \nabla U) \nonumber\\
&+& 24\frac{G}{c^2} a_2^2 \V(\nabla\rho \cdot \nabla R) - \frac{192\pi G^2}{c^4} a_2q_1 \V(|\nabla\rho|^2)
+ \frac{2G}{c^4} \V(\rho\Pi) + \frac{4G}{c^4} \V(\rho v^2) + \frac{6G}{c^4} \V(p) \nonumber\\
&-& \frac{1}{6\pi c^2} X(UR) + \frac{1}{4\pi}\left( a_2+\frac{a_3}{2a_2} \right)X(R^2) +
\frac{4G}{3c^4} X(\rho U) \nonumber\\
&-& \frac{G}{6c^2}\left( 16a_2+20q_1+8\frac{q_2}{a_2} \right)X(\rho R)
+ \frac{16\pi G^2}{3c^4} q_1\left( 4-\frac{q_1}{a_2} \right)X(\rho^2) \nonumber \\
&-& \frac{8G}{3c^4}\left[ a_2-q_1\left(2-\frac{q_1}{a_2}\right)\right]X(\nabla\rho \cdot \nabla U)
+ \frac{8G}{c^2} a_2\left(a_2-q_1\right)X(\nabla\rho \cdot \nabla R) \nonumber\\
&-& \frac{64\pi G^2}{c^4} q_1\left(a_2-q_1\right)X(|\nabla\rho|^2) - \frac{2G}{c^4} X(p)
+\frac{2G}{3c^4} X(\rho\Pi) - \frac{1}{c^4}\sqrt{\frac{2}{3} a_2}\left( 1 - \frac{q_1}{a_2} \right)\hat\chi_{,00},
\nonumber
\end{eqnarray}
\end{widetext}
where, for brevity, $R$ denotes $R^{(2)}$, and the Poisson and Yukawa potentials $\V$ and $X$, respectively, are defined by
\begin{eqnarray}
\V(Q) &=& \int \frac{Q(t,\y)}{|\x-\y|} d^3y, \\ \nonumber
X(Q) &=& \int Q(t,\y) \frac{e^{-m|\x-\y|}}{|\x-\y|} d^3y,
\end{eqnarray}
while the potential $\hat\chi$ is given by
\begin{equation}
\hat\chi = G\int \rho(t,\y)e^{-m|\x-\y|}d^3y.
\end{equation}
In the case of pure $f(R)$ gravity Eq. (\ref{h004}) differs
from the solution for $h_{00}^{(4)}$ found in Ref. \cite{Cl} for some coefficients of order of unity.

The expression for $h_{00}^{(4)}$ is not in the usual PPN form, since it contains both the time derivatives
$U_{,00}$, $R_{,00}$, $\rho_{,00}$ and $\hat\chi_{,00}$, and terms depending on the gradients
$\nabla U$, $\nabla R$ and $\nabla\rho$. In the following section, these will be eliminated by means of suitable gauge transformations, thus yielding a more adequate PPNY form for the metric.

\subsection{Gauge transformation}\label{sec:gauge}
So far, the metric has been computed in the gauge specified by conditions Eqs. (\ref{gauge-1}), which are convenient in the PPN framework \cite{Wi}. However, the solution Eq. (\ref{hij-2}) for the metric perturbation $h_{ij}$ at second order is not diagonal, hence it is not in the standard post-Newtonian gauge. Moreover, we recall that the metric perturbation $h_{00}$ at fourth order also contains terms that do not appear in the standard post-Newtonian approximation.

To correct this, we follow Ref. \cite{Cl} and make a further gauge transformation
\begin{equation}x^\mu \rightarrow x^\mu + \xi^\mu,
\end{equation}
so that the metric perturbation transforms as
\begin{equation}
h_{\mu\nu} \rightarrow h_{\mu\nu} - \nabla_\nu \xi_\mu - \nabla_\mu\xi_\nu + O(\xi^2) .
\end{equation}
Adopting the form
\begin{eqnarray}
\xi_0 &=& \frac{6}{c^3} \left(a_2-q_1\right)U_{,0} - 18\frac{a_2^2}{c} R_{,0} + \\ \nonumber &&
\frac{144\pi G}{c^3} a_2q_1\rho,0 - \frac{1}{c^3}\sqrt{\frac{a_2}{6}}\left(1-\frac{q_1}{a_2}\right)\hat\chi_{,0}, \\ \nonumber
\xi_i &=& \frac{4}{c^2}\left(a_2-q_1\right)U_{,i} - 12a_2^2 R_{,i} + \frac{96 \pi G}{c^2} a_2q_1\rho_{,i},
\end{eqnarray}
the metric perturbation $h_{\mu\nu}$ transforms into a diagonal expression, with no time derivatives and some of the gradient terms in $h_{00}^{(4)}$ are gauged away:
\begin{widetext}
\begin{eqnarray}
h_{ij}^{(2)} &\rightarrow& h_{ij}^{(2)} - \frac{8}{c^2}\left(a_2-q_1\right)U_{,ij} + 24a_2^2R_{,ij} -
\frac{192\pi G}{c^2} a_2q_1\rho_{,ij}, \\ \nonumber
h_{0i}^{(3)} &\rightarrow& h_{0i}^{(3)} - \frac{10}{c^3}\left(a_2-q_1\right)U_{,0i} + 30\frac{a_2^2}{c} R_{,0i} -
\frac{240\pi G}{c^3} a_2q_1\rho_{,0i} + \frac{1}{c^3}\sqrt{\frac{a_2}{6}}\left(1-\frac{q_1}{a_2}\right)
\hat\chi_{,0i}, \\ \nonumber
h_{00}^{(4)} &\rightarrow& h_{00}^{(4)} + 36\frac{a_2^2}{c^2} R_{,00} - \frac{12}{c^4}\left(a_2-q_1\right)U_{,00}
-\frac{288\pi G}{c^4} a_2q_1\rho_{,00} + \frac{1}{c^4}\sqrt{\frac{2}{3} a_2}\left( 1 - \frac{q_1}{a_2} \right)
\hat\chi_{,00} \\ \nonumber
&-& \frac{8}{c^4}\left(a_2-q_1\right)|\nabla U|^2 + 24a_2^3|\nabla R|^2 +
\frac{1536\pi^2G^2}{c^4} a_2q_1^2|\nabla\rho|^2 +8\frac{a_2}{c^2}\left(2a_2+q_1\right)\nabla U \cdot \nabla R \\ \nonumber
&-& \frac{64\pi G}{c^4} q_1\left(2a_2+q_1\right) \nabla\rho \cdot \nabla U -
\frac{384\pi G}{c^2} a_2^2q_1 \nabla\rho \cdot \nabla R~~.
\end{eqnarray}
\end{widetext}
Using the continuity Eq. (\ref{contin-eq}), the quantity $\hat\chi_{,0i}$ appearing in the transformation law
for $h_{0i}^{(3)}$ is given by
\begin{equation}
\hat\chi_{,0i} = \frac{1}{\sqrt{6a_2}}\left(GX(\rho v_i) -Y_i - \frac{1}{\sqrt{6a_2}}Z_i \right) ,
\end{equation}
where the potentials $Y_i$ and $Z_i$ are defined as
\begin{equation}
Y_i = G \int \frac{\rho(t,\y)[\v(t,\y)\cdot(\x-\y)](x-y)_i}{|\x-\y|^3} e^{-m|\x-\y|}d^3 y,
\end{equation}
and
\begin{equation}
Z_i = G \int \frac{\rho(t,\y)[\v(t,\y)\cdot(\x-\y)](x-y)_i}{|\x-\y|^2} e^{-m|\x-\y|}d^3 y.
\end{equation}
Collecting the results from Sections \ref{sec:nonrel-limit} and \ref{sec:PPNY},
the form of the metric after the gauge transformations is
\begin{widetext}
\begin{eqnarray}\label{g00}
g_{00} &=& -1 + 2\left(\frac{U}{c^2} + a_2R - \frac{8\pi G}{c^2} q_1\rho \right) -
\frac{2}{c^4} U^2 - 2a_2^2R^2 -4\frac{a_2}{c^2} UR +
\frac{32\pi G}{c^2} q_1 \rho \left( \frac{U}{c^2} + a_2 R \right) \\
&- & \frac{128\pi^2G^2}{c^4} q_1^2\rho^2
-\frac{a_2}{3\pi} \V(R^2) + \frac{4G}{c^4} \V(\rho U) -  \frac{2G}{c^2}\left(\frac{8}{3} a_2-5q_1\right)\V(\rho R) + \frac{64\pi G^2}{c^4} q_1\V(\rho^2) \nonumber\\ &- & \frac{8G}{c^4}\left(a_2-q_1\right)\V(\nabla\rho \cdot \nabla U) + 24\frac{G}{c^2} a_2^2 \V(\nabla\rho \cdot \nabla R) - \frac{192\pi G^2}{c^4} a_2q_1 \V(|\nabla\rho|^2)
+ \frac{2G}{c^4} \V(\rho\Pi) + \frac{4G}{c^4} \V(\rho v^2) \nonumber\\
&+& \frac{6G}{c^4} \V(p) - \frac{1}{6\pi c^2} X(UR) + \frac{1}{4\pi}\left( a_2+\frac{a_3}{2a_2} \right)X(R^2) +
\frac{4G}{3c^4} X(\rho U) - \frac{G}{6c^2}\left( 16a_2+20q_1+8\frac{q_2}{a_2} \right)X(\rho R)\nonumber\\
&+& \frac{16\pi G^2}{3c^4} q_1\left( 4-\frac{q_1}{a_2} \right)X(\rho^2) - \frac{8G}{3c^4}\left[ a_2-q_1\left(2-\frac{q_1}{a_2}\right)\right]X(\nabla\rho \cdot \nabla U)
+ \frac{8G}{c^2} a_2\left(a_2-q_1\right)X(\nabla\rho \cdot \nabla R) \nonumber\\ \nonumber
&-& \frac{64\pi G^2}{c^4} q_1\left(a_2-q_1\right)X(|\nabla\rho|^2) - \frac{2G}{c^4} X(p)
+\frac{2G}{3c^4} X(\rho\Pi), \end{eqnarray}
\end{widetext}
\begin{widetext}
\begin{eqnarray}
g_{0i} &=& -\frac{7}{2c^3} V_i - \frac{1}{2c^3} W_i + \frac{1}{6c^3}\left(1-\frac{q_1}{a_2}\right)\left[ GX(\rho v_i) -
Y_i - \frac{1}{\sqrt{6a_2}} Z_i\right], \\~\nonumber\\
g_{ij} &=& \left[ 1 + 2\left( \frac{U}{c^2} - a_2R + \frac{8\pi G}{c^2} q_1\rho \right) \right] \delta_{ij}.
\end{eqnarray}
\end{widetext}
The spatial part of the metric $g_{ij}$ is now diagonal, as in the standard post-Newtonian gauge.
However, although time derivatives have been eliminated from $g_{00}$, the latter is not yet in the usual PPN form, since it
contains contributions with the potentials $\V$ and $X$ depending on the gradient terms
$\nabla\rho \cdot \nabla U$, $\nabla\rho \cdot \nabla R$ and $|\nabla\rho|^2$.

Once again, following Ref. \cite{Cl}, we transform such potentials into expressions without gradient terms: in this section we show, for instance, how the gradient terms can be eliminated from the contribution $X(\nabla\rho \cdot \nabla U)$. The complete computations are given in Appendix B,
where all the undesired terms are transformed into expressions free of gradient terms, resorting instead to eleven new potentials $\psi_i$ ($i=0,...,10$). We have
\begin{equation}
X(\nabla\rho \cdot \nabla U) = \int \frac{\nabla\rho(t,\y) \cdot \nabla U(t,\y)}{|\x-\y|}
e^{-m|\x-\y|} d^3y.
\end{equation}
Introducing the vector field
\begin{equation}
\mathbf{A}(t,\x,\y) = \frac{e^{-m|\x-\y|}}{|\x-\y|} \nabla U(t,\y),
\end{equation}
and using the divergence theorem and the boundary conditions (\ref{boundary-cond}) yields
\begin{eqnarray}
X(\nabla\rho \cdot \nabla U) &=& \int \nabla\rho(t,\y) \cdot \mathbf{A}(t,\x,\y) d^3y \\ \nonumber & =&
-\int \rho(t,\y) \nabla_{y} \cdot \mathbf{A}(t,\x,\y) d^3y,
\end{eqnarray}
where the operator $\nabla_{y}$ denotes the divergence with respect to the coordinates $\y$.
The evaluation of the divergence of the vector field $\mathbf{A}$, using the Poisson equation
$\nabla^2_{y}U(t,\y)=-4\pi G\rho(t,\y)$, yields
\begin{eqnarray}
&& \nabla_{y} \cdot \mathbf{A}(t,\x,\y) = -4\pi G\rho(t,\y)\frac{e^{-m|\x-\y|}}{|\x-\y|} \\ \nonumber && -
Gm\int \frac{\rho(t,\z)(\x-\y)\cdot(\y-\z)}{|\x-\y|^2|\y-\z|^3}
e^{-m|\x-\y|} d^3z \\ \nonumber
&-& G\int \frac{\rho(t,\z)(\x-\y)\cdot(\y-\z)}{|\x-\y|^3|\y-\z|^3}
e^{-m|\x-\y|} d^3z,
\end{eqnarray}
from which we obtain the following expression with the gradient terms expunged:
\begin{equation}
X(\nabla\rho \cdot \nabla U) = 4\pi G X(\rho^2) + \frac{\psi_4}{G\sqrt{6a_2}} + \frac{\psi_5}{G},
\end{equation}
where the potentials $\psi_4$ and $\psi_5$ are defined by
\begin{eqnarray}
&&\psi_4 = G^2 \times \\ \nonumber && \int \frac{\rho(t,\y)\rho(t,\z)(\x-\y)\cdot(\y-\z)}{|\x-\y|^2|\y-\z|^3}
e^{-m|\x-\y|} d^3yd^3z, \\ \nonumber
&& \psi_5 = G^2 \times \\ \nonumber && \int \frac{\rho(t,\y)\rho(t,\z)(\x-\y)\cdot(\y-\z)}{|\x-\y|^3|\y-\z|^3}
e^{-m|\x-\y|} d^3yd^3z.
\end{eqnarray}
%

\subsection{PPNY metric}

In this section we denote by $\Y$ the Yukawa potential generated by a distribution of masses with density $\rho$:
\begin{equation}\label{Ypotential}
\Y = G\int \rho(t,\y) \frac{e^{-m|\x-\y|}}{|\x-\y|} d^3y,
\end{equation}
so that Eq. (\ref{R2-solution}) can be written as
\begin{equation}\label{R2-sol-rewritt}
R^{(2)} = \frac{1-\theta}{3c^2a_2} \Y + \frac{8\pi G}{c^2} \theta\rho,
\end{equation}
where we define the dimensionless parameter $ \theta = q_1\slash a_2$.
Next we introduce the standard PPN potentials \cite{Wi}, constructed with the Poisson kernel:
\begin{eqnarray}\label{PPN-potentials}
\Phi_1 = G\V(\rho v^2)~~ &,& ~~ \Phi_2 = G\V(\rho U), \\ \nonumber ~~ \Phi_3 = G\V(\rho\Pi)~~ &,& ~~ \Phi_4 = G\V(p),
\end{eqnarray}
and the analogous potentials, constructed with the Yukawa kernel, which are characteristic of the NMC gravity model:
\begin{equation}\label{NMC-potentials}
\Sigma_2 = G X(\rho U), ~~ \Sigma_3 = G X(\rho\Pi), ~~ \Sigma_4 = G X(p).
\end{equation}
Moreover, we introduce the following new potentials
\begin{equation}\label{new-potentials}
\Phi_5 = G^2\V(\rho^2), ~~ \Sigma_5 = G^2 X(\rho^2), ~~ \Theta_i = G X(\rho v_i).
\end{equation}
Substituting into Eq. (\ref{g00}) the expression for the curvature (\ref{R2-sol-rewritt}), the potentials given in Eqs.
(\ref{PPN-potentials}), (\ref{NMC-potentials}) and (\ref{new-potentials}),
and the potentials $\psi_i$ ($i=0,...,10$) given in Appendix B, we obtain the final expression for the metric tensor,
\begin{widetext}
\begin{eqnarray}\label{g00-final}
g_{00} &=& -1 + 2\frac{U}{c^2} + (1-\theta)\frac{2}{3c^2} \Y - \frac{2}{c^4} U^2 +
\frac{2}{c^4}\left(2\Phi_1+2\Phi_2+\Phi_3+3\Phi_4\right) \\
&+& \frac{2}{c^4}\left[ \frac{2}{3}(1-\theta)\Sigma_2 +\frac{1}{3} \Sigma_3 -\Sigma_4 \right]
+ \frac{16\pi}{3c^4} \theta\left(4a_2+11q_1\right)\Phi_5 \nonumber\\
&+& \frac{8\pi}{c^4} \theta \left( -2q_1+\frac{a_3q_1}{a_2^2} - \frac{4}{3} \frac{q_2}{a_2} \right) \Sigma_5
+ \frac{(1-\theta)}{c^4}\left\{ -\frac{2}{9}(1-\theta)\Y^2 - \frac{4}{3}U\Y
-\frac{1}{18\pi a_2} X(U\Y) \right. \nonumber\\
&-& \frac{1}{9\pi} \frac{(1-\theta)}{a_2}\left[\frac{1}{3} \V(\Y^2) - \frac{1}{4}\left(1+\frac{a_3}{2a_2^2}\right)
X(\Y^2) \right] - \frac{14}{9}(2-\theta)\psi_0 -8a_2\psi_1 \nonumber \\
&+& 2\sqrt{\frac{a_2}{3}} \psi_2 + 8a_2\psi_3 -
\frac{4}{3}(1-\theta)a_2\left[ \sqrt{\frac{2}{3a_2}} \psi_4 +2\psi_5
-2\psi_6 -\sqrt{\frac{2}{3a_2}}\left(\psi_7+\psi_8\right) - \frac{1}{3a_2} \psi_9 \right] \nonumber\\
&+& \left. \frac{2}{3a_2}\left( -2a_2 + q_1 +\frac{a_3q_1}{a_2^2} - \frac{2}{3} \frac{q_2}{a_2} \right)\psi_{10}
\right\}, \nonumber\\
g_{0i} &=& -\frac{7}{2c^3} V_i - \frac{1}{2c^3} W_i + \frac{1}{6c^3}(1-\theta) \left( \Theta_i -
Y_i - \frac{1}{\sqrt{6a_2}} Z_i\right), \nonumber\\
\nonumber\\
g_{ij} &=& \left[ 1 + 2\frac{U}{c^2} - (1-\theta)\frac{2}{3c^2} \Y \right] \delta_{ij}.
\nonumber
\end{eqnarray}
\end{widetext}
%

\section{Static, spherically symmetric metric around a body with uniform density}\label{sec:spheric-metric}

In this section we give the expression for the PPNY metric in vacuum, around a spherical
body of radius $R_S$ and with a static, uniform mass density: hence, we assume
$\rho(t,x)=const.$ inside the body and $v=0$. This is a simple model which allow us to achieve an
explicit expression for the metric amenable for computation of orbits around a body
(either the Sun or a planet) in the Solar System.

Note that such a mass density does not satisfy the boundary conditions
Eq. (\ref{boundary-cond}) at the surface of the body. Nevertheless, in order to satisfy such boundary
conditions, we may model the mass density of the body with a constant value in an interior region
and a sharp transition in a thin layer close to the surface. When the thickness of the layer tends
to zero, the various potentials appearing in the PPNY metric converge to the potentials corresponding
to a uniform density model, since such potentials depend only on the density $\rho$ and not on spatial
derivatives of $\rho$. Hence, the uniform density model is an approximation (limit case) of a density model
with a thin layer. In what follows, we set the origin of the spatial coordinates at the center of the spherical body and set
$r=|x|$.

\subsection{Effective mass}

In order to find the expression for the metric, we first observe that all the potentials
in the $g_{00}$ coefficient of the
PPNY metric which involve the Poisson integral --- {\it i.e.} the potentials of the type $\V(Q)$
with the exception of $\V(\Y^2)$, under our assumptions on the density $\rho$ ---
are proportional to $1\slash r$ outside of the body, whenever $r>R_S$.
The potential $\V(\Y^2)$ has to be decomposed into the sum of two
potentials $\V_1(\Y^2)+\V_2(\Y^2)$, where $\V_1$ is proportional to $1\slash r$ in vacuum, while $\V_2$ contains
other functions of $r$ (see Appendix C).

Hence, we can take into account the potentials proportional to $1\slash r$, for $r>R_S$, absorbing such
contributions  in the effective mass $M_S$ of the body, defined as follows:
\begin{eqnarray}
\frac{GM_S}{r} &=& U + \frac{1}{c^2}\left(2\Phi_1+2\Phi_2+\Phi_3+3\Phi_4\right) \nonumber \\ \nonumber
&+&\frac{8}{3} \frac{\pi}{c^2} \theta\left(4a_2+11q_1\right)\Phi_5 \\ \nonumber
&-& \frac{(1-\theta)}{2c^2}\left[ \frac{1}{27\pi} \frac{(1-\theta)}{a_2} \V_1(\Y^2) \right. \\
&+& \left. \frac{14}{9}(2-\theta)\psi_0 \right],
\end{eqnarray}
where $\psi_0 = G\V(\rho\Y)$.

In all other potentials of $g_{00}$ we can replace the Newtonian mass with the effective mass,
{\it i.e.} $\int\rho(\x)d^3x \rightarrow M_S$, keeping the accuracy of the $O(1\slash c^4)$ approximation.

\subsection{Yukawa potential}\label{subsec:Yukawa}

In the case of pure $f(R)$ gravity, i.e., $q_1=q_2=0$, it turns out that most of the terms in $g_{00}$
are negligible because of exponential suppression \cite{Cl}: this reflects the requirement for a short ranged Yukawa interaction, so as to make it compatible with observations \cite{CSE}.
Conversely, in NMC gravity the Yukawa interaction can be long ranged, as it has been shown in Ref. \cite{CPM},
so that terms which are not exponentially suppressed arise in $g_{00}$, which in principle allow to constrain the theory
by means of Solar System experiments.

We now observe that, assuming a constant density $\rho$, all the potentials in the $g_{00}$ coefficient of the
PPNY metric which involve the Yukawa integral, hence the potentials of the type $X(Q)$
with the exception of $X(U\Y)$ and $X(\Y^2)$, are proportional to $\exp(-r/\lambda)\slash r$ outside of the body, $r>R_S$ (where $\lambda = 1\slash m =\sqrt{6a_2}$). The potentials $X(U\Y)$ and $X(\Y^2)$, evaluated in vacuum, contain both terms proportional to
$\exp(-r/\lambda)\slash r$ and other functions of $r$ (see Appendix C).

We can take into account the potentials proportional to $\exp(-r/\lambda)\slash r$, for $r>R_S$,
absorbing such contributions in the effective strength $\alpha$ of a Yukawa potential, which yields the following
contribution to $g_{00}$:
\begin{equation}
\frac{2}{c^2} GM_S \alpha  \frac{e^{-r\slash\lambda}}{r}.
\end{equation}
The expressions of all the potentials of the type $X(Q)$ appearing in $g_{00}$ are listed in Appendix C.
Assembling such expressions, it turns out that the effective strength $\alpha$ is a function of the following
four dimensionless quantities built with the parameters of the considered NMC model:
\begin{equation}\label{dimensionlessparameters}
\theta=\frac{q_1}{a_2} ,~~ \mu=\frac{a_3}{a_2^2} , ~~ \nu=\frac{q_2}{a_2^2} , ~~ \frac{R_S}{\lambda}.
\end{equation}
We assume that the range of the Yukawa potential satisfies the condition $\lambda \gg R_S$, and
we expand the potentials in power series of $R_S/\lambda$. Again, we remark that in the case
of pure $f(R)$ gravity, if $R_S$ is the radius either of the Sun or of the Earth, then the condition
$\lambda \gg R_S$ is not compatible with Solar System observations \cite{CSE,NJ}.

Keeping only powers $(R_S/\lambda)^n$ with $n\leq 2$, and using the results given in Appendix C, it follows that $\alpha$ can be decomposed into a zeroth-order and a first order contribution on $1/c^2$:
\begin{eqnarray}
\label{defalpha}\alpha &=& \alpha_0 + \frac{GM_S}{c^2R_S}\alpha_1 , \\ \nonumber \alpha_0 &=& \frac{1}{3}(1-\theta)\left[ 1 + \frac{1}{10}\left(\frac{R_S}{\lambda}\right)^2 \right] \\ \nonumber
\alpha_1 &=&\sum_{n=-2}^2 A_n \left(\frac{R_S}{\lambda}\right)^n,
\end{eqnarray}
with the coefficients $A_i$ given by
\begin{eqnarray}
A_{-2} &=& -\theta \left[ \theta \left( 1 - \frac{\mu}{2} \right) + \frac{2}{3} \nu \right], \\ \nonumber A_{-1}&=&0 ,\\ \nonumber A_0 &=& -\frac{8}{45} \bigg[ 1 + \frac{3}{2}\nu - \theta \left( 2 + \frac{9}{4}\mu + \frac{9}{8} \nu \right) +\\ \nonumber && \frac{1}{32}\theta^2 (86+63\mu) \bigg], \\ \nonumber A_1 &=& -\frac{1}{108}(1-\theta) \bigg\{ -68+32\theta+36\theta\mu-24\nu + \\ \nonumber && 36 \mbox{Ei}\left( -2 \frac{R_S}{\lambda}\right) + 9 (1-\theta) (2 + \mu) \times \\ \nonumber &&  \left[ \mbox{Ei}\left(- \frac{R_S}{\lambda}\right) - \mbox{Ei}\left( - 3\frac{R_S}{\lambda}\right) \right]  \bigg\} ,\\ \nonumber A_2 &=& -\frac{89}{175} \bigg[ 1- \frac{85}{534}\mu + \frac{220}{801}\nu - \\ \nonumber && \frac{1}{89}\theta \left( 93 + \frac{25}{3}\mu + \frac{865}{36}\nu \right) + \\ \nonumber && \frac{1}{8} \theta^2 \left( \frac{57}{89} + \frac{355}{178}\mu \right) \bigg].
\end{eqnarray}
where $\mbox{Ei}(x)$ denotes the exponential integral function:
\begin{equation}
\label{Eifunction}
\mbox{Ei}(x) = -\int_{-x}^{+\infty}\frac{e^{-t}}{t} dt .
\end{equation}
We conclude this section observing that, if the condition $\lambda \gg R_S$ is not satisfied, then most
of the terms in $g_{00}$ are exponentially suppressed if $r>R_S$, so that they become quickly negligible by
increasing $r$ outside of the body. The only potentials which are not present
in GR and are not exponentially suppressed (see also Ref. \cite{Cl}) are $\psi_1$
(if either $\lambda \ll R_S$ or $\lambda \approx R_S$) and $\psi_2,\psi_3$ (if $\lambda \approx R_S$).
However, for a static, spherically symmetric (not necessarily uniform)
mass density $\rho=\rho(r)$, we find that such potentials vanish identically for $r>R_S$.

\subsection{Further potentials}

Using the expression Eq. (\ref{g00-final}) for the PPNY metric and the results given in Appendix C,
it turns out that the coefficient $g_{00}$ contains the following combination of functions of $r$:
\begin{eqnarray}
& &-\frac{2}{c^4}\left(\frac{GM_S}{r}\right)^2\left( 1 + \beta_1 e^{-r/\lambda} + \beta_2 e^{-2r/\lambda} \right) \nonumber \\
& &-\frac{2}{c^2} \frac{GM_S}{r}\left(\frac{GM_S}{c^2R_S}\right)\sum_{i=1}^3 \zeta_i F_i(r),
\end{eqnarray}
with the coefficients
\begin{eqnarray}\label{defcoefficients}
\beta_1 &=& \frac{2}{3}(1-\theta)\left[ 1+\frac{1}{10}\left(\frac{R_S}{\lambda}\right)^2 \right], \\ \nonumber
\beta_2 &=& \frac{1}{9}(1-\theta)^2\left[ 1+\frac{1}{5}\left(\frac{R_S}{\lambda}\right)^2 \right], \\ \nonumber
\zeta_1 &=& \frac{1}{3} (1-\theta)\frac{R_S}{\lambda}, \\ \nonumber
\zeta_2 &=& \frac{2}{9} (1-\theta)^2\frac{R_S}{\lambda}, \\ \nonumber
\zeta_3 &=& -\frac{1}{6} (1-\theta)^2\left(1+\frac{\mu}{2}\right)\frac{R_S}{\lambda},
\end{eqnarray}
and the three functions $F_i(r)$ given by
\begin{eqnarray}
F_1(r) &=& e^{-r/\lambda}\ln\left(\frac{r}{R_S}\right) - e^{r/\lambda}\mbox{Ei}\left(-\frac{2r}{\lambda}\right),\nonumber \\
F_2(r) &=& e^{-2r/\lambda} + 2 \frac{r}{\lambda} \mbox{Ei}\left(-\frac{2r}{\lambda}\right), \label{Ffunctions} \\ \nonumber
F_3(r) &=& e^{-r/\lambda}\mbox{Ei}\left(-\frac{r}{\lambda}\right) - e^{r/\lambda}\mbox{Ei}\left(-\frac{3r}{\lambda}\right).
\end{eqnarray}
%
\subsection{PPNY metric around the spherical body}

Collecting the results of the previous sections, we find the expression for the metric tensor,
\begin{widetext}
\begin{eqnarray}\label{g00-spherical}
g_{00} &=& -1 + 2 \frac{GM_S}{rc^2}\left( 1 + \alpha e^{-r/\lambda} \right)\\ \nonumber
&-& \frac{2}{r}\left(\frac{GM_S}{c^2}\right)^2\left( \frac{1}{R_S}\Big[ \zeta_1  F_1(r) +
\zeta_2  F_2(r) + \zeta_3  F_3(r) \Big] + \frac{1}{r}\left[ 1 + \beta_1 e^{-r/\lambda} + \beta_2 e^{-2r/\lambda} \right] \right), \\ \nonumber \\ \nonumber
g_{0i} &=& 0, \\ \nonumber \\ \nonumber
g_{ij} &=& \left[ 1 + 2 \frac{GM_S}{rc^2}\left( 1 - \alpha e^{-r/\lambda} \right) \right] \delta_{ij}.
\nonumber
\end{eqnarray}
\end{widetext}
%
\section{Perihelion Precession}

In this section we use the previously obtained expression for the PPNY metric, Eq. (\ref{g00-spherical}), to assess the impact of the NMC gravity model on the precession of the perihelion of closed orbits.

Alternatively, a coordinate transformation to the usual non-isotropic Schwarzschild frame could be performed (see Refs. \cite{Hu,Adkins,Schmidt}). As shown in Ref. \cite{JiangLin} for the case of General Relativity, both approaches naturally lead to the same result, highlighting the general covariance of the theory, maintained by the NMC model here considered.

The action for a point particle with mass $m$ is given by
\begin{equation}
S= mc\int d\tau [1+f^2(R)] \sqrt{ -g_{\mu\nu} \frac{dx^\mu}{d\tau}\frac{dx^\nu}{d\tau}},
\end{equation}
where $\tau$ is an affine parameter (which, for the case of timelike geodesics, can be identified with the proper time). This is invariant for reparameterisations of the form $x^\mu (\tau) \to x^\mu(\tau) + \delta x^\mu$, so that variations with respect to $\delta x^\mu$ yield the equations of motion \cite{newtlimit},
\begin{equation}\label{geodesic}
\frac{d^2 x^\alpha}{ds^2} + \Gamma^\alpha_{\mu\nu} \frac{dx^\mu}{ds} \frac{dx^\nu}{ds} = \frac{f^2_R(R)}{ 1+f^2(R)} g^{\alpha\beta} R_{,\beta},
\end{equation}
\noindent clearly showing that the NMC gravity model under scrutiny leads to a deviation from geodesic motion \cite{BBHL,Sotiriou1}.

Naturally, we are considering the test body to travel outside the central mass. However, this does not imply that the scalar curvature vanishes, as Eq. (\ref{R2-sol-rewritt}) shows to order $O(1/c^2)$. Furthermore, one must consider the contribution to this term of both the Yukawa potential given in Eq. (\ref{Ypotential}) generated by both the central body as well as the test body itself, $ \Y = \Y_S + \Y_B$ --- thus giving rise to the possibility of a self-acceleration.

If the test body has inner structure ({\it e.g.} a density $\rho_B(t,x)$), this will further complicate the computation of the additional force arising from the non-conservation of the energy-momentum tensor depicted on the {\it r.h.s.} of the above. As such, we consider that the test body is homogeneous and static, $\rho_B(t,x) =const.$, consistent with the approximation considered in the previous section for the central body itself (for a thorough discussion of the effect of the inner structure on the non-geodesic motion induced by a NMC model, see Ref. \cite{puetzobukiorio}).

We must still consider the effect of the Yukawa potentials arising both from the central body as well as the test body. For this, we resort to Eq. (\ref{Ypotentialhomogeneous}) of Appendix C, where this quantity is computed assuming a homogeneous density $\rho$,
\begin{equation}
\Y = GM_S \frac{e^{-r/\lambda}}{r}  \left[ 1 + \frac{1}{10}\left(\frac{R_S}{\lambda}\right)^2 \right].
\end{equation}
Anticipating the comparison with the observed precession of the perihelion of Mercury, we may compute the proportion between both contributions,
\begin{equation}
\frac{\Y_B}{\Y_S} \approx \frac{M_B }{M_S}\frac{L}{r},
\end{equation}
\noindent where condition $\lambda \gg R_S $ was considered, $ L \sim 55 \times 10^9$ m is the characteristic distance from Mercury to the Sun and $r$ is the distance to the centre of the planet.
Since $M_B \sim 1.7\times 10^{-7}M_S$, we find that the Yukawa potential created by Mercury itself is only dominant up to a distance to its centre $r \lesssim 10$ km $\approx 0.3\%$ of its radius. Thus, we conclude that we may safely disregard the self-acceleration of Mercury due to the Yukawa potential it generates, and focus solely on the contribution of the Sun, $\Y \sim \Y_S$.

In order to compute perturbations to the Newtonian orbit, it is useful to write the equations of motion in the form
\begin{equation}
\frac{d^2 x^i}{d t^2} = - \left( \Gamma^i_{\alpha\beta} - \Gamma^0_{\alpha\beta}\,
\frac{\dot x^i}{c} \right)\dot x^\alpha \dot x^\beta + \delta_r^i N(r),
\end{equation}
where dot denotes time derivative and $N(r)$ is the additional potential due to the non-conservation of the energy-momentum tensor,
\begin{eqnarray}N(r) = c^2\frac{g_{00}}{g_{jj}}\frac{q_1 + 2 q_2R}{ 1+q_1 R + q_2R^2}R'(r),
\end{eqnarray}
and the factor $g_{00}$ is due to the transformation $ds \to dt$.
To the desired order $O(1/c^4)$ on the metric $g_{\mu\nu}$ and the scalar curvature $R$, we have
\begin{eqnarray} N(r) &=& -\frac{c^2}{6}\theta \lambda^2\Bigg[  {R^{(4)}}'(r) + \bigg( 1 - \frac{v^2}{c^2} - h_{00}^{(2)} - h_{jj}^{(2)} \nonumber \\ \label{Nexpression} &+&\frac{ \lambda^2}{6} R^{(2)}\left[2\frac{\nu}{\theta} - \theta \right] \bigg) {R^{(2)}}'(r) \Bigg].\end{eqnarray}

Here and in the sequel the prime denotes derivative with respect to $r$.
Using Eq. (\ref{R2-sol-rewritt}), the potentials defined in Appendix C and the definitions Eq. (\ref{defalpha}), we can write the scalar curvature to the required order,
\begin{eqnarray}
R^{(2)} &=& 2\frac{1-\theta}{c^2\lambda^2} \Y = \frac{6\alpha_0}{\lambda^2 } \frac{GM_S}{r c^2} e^{-r/\lambda} , \\ \nonumber R^{(4)} &=& \frac{4}{c^4} Y_0' \left[(1-\theta)U^\prime -3 Y_0^\prime \right] \\ \nonumber &+&\frac{6}{c^4\lambda^2}\bigg[ c^2 Y_1 + \left( 1-\frac{3}{2}\mu \right)Y_0^2 \\ \nonumber &-& \frac{(GM_S)^2}{R_S r}\left( \zeta_1 F_1 + \zeta_3 F_3 \right)\bigg],
\end{eqnarray}
where we define the Yukawa contributions
\begin{eqnarray}
Y_0(r) &=& \alpha_0 \frac{GM_S}{r} e^{-r/\lambda}, \\ \nonumber Y_1(r) &=& \alpha_1 \frac{(GM_S)^2}{R_Sr c^2} e^{-r/\lambda},\\ \nonumber Y(r) &=& Y_0(r) + Y_1(r) =  \alpha \frac{GM_S}{r} e^{-r\slash\lambda}.
\end{eqnarray}
We thus obtain the expression below,
\begin{eqnarray}
&& N(r) = \\ \nonumber && \frac{\theta}{c^2}\Bigg[ \left[4U + v^2 - c^2 -
\left(2-\theta-3\mu+2\frac{\nu}{\theta}\right)Y_0\right]Y_0^\prime \\ \nonumber
&&- c^2 Y_1^\prime + 4\lambda^2 Y_0^\prime Y_0^{\prime\prime} - \frac{2}{3}\lambda^2(1-\theta)
\left( U^{\prime\prime}Y_0^\prime + U^\prime Y_0^{\prime\prime} \right) \\ \nonumber
&& + \frac{(GM_S)^2}{R_S r}\left[ \zeta_1\left( F_1^\prime - \frac{F_1}{r} \right) +
\zeta_3\left(F_3^\prime - \frac{F_3}{r} \right) \right]\Bigg] ,
\end{eqnarray}
valid to order $O(1/c^4)$.

In the following, we set
\begin{eqnarray}
F(r) =
-\frac{(GM_S)^2}{r}&\Bigg[&\frac{1}{r}\left( 1 + \beta_1 e^{-r/\lambda} + \beta_2 e^{-2r/\lambda} \right)  \nonumber \\ \label{Feq}
&+& \frac{1}{R_S}\sum_{i=1}^3 \zeta_i F_i(r) \Bigg].
\end{eqnarray}
Using the metric Eq. (\ref{g00-spherical}), the equations of motion yield
\begin{equation}\label{geodesics}
\frac{d \mathbf{v}}{d t} = -\frac{GM_S \mathbf{r}}{ r^3} + \mathbf{\Delta},
\end{equation}
with the perturbative force
\begin{eqnarray}\mathbf{\Delta} &= &\Bigg[ Y^\prime - \frac{2}{c^2}\left( U-Y\right) (U^\prime + Y^\prime) + \frac{v^2}{c^2} (U^\prime - Y^\prime)\nonumber \\ \label{perturbativeforce}
&&+ \frac{F'}{c^2} + N(r)\Bigg] \frac{\mathbf{r}}{r} - 4U^\prime\frac{ \dot{r} }{c^2}\mathbf{v}  \equiv \\ \nonumber && \Delta_r \frac{\mathbf{r}}{r} + \Delta_v \frac{\mathbf{v}}{v},
\end{eqnarray}
where $v = |\mathbf{v}|$, and $\Delta_r$, $\Delta_v$ are defined implicitly.

To compute the precession of the perihelion, we follow Refs. \cite{Weinberg2,JiangLin} and begin by recalling that, in Newtonian Mechanics, orbits are ellipses (with perihelion at an angle $\phi = \phi_P$), described by
\begin{equation}
r(\phi) = \frac{L}{1+e \cos(\phi - \phi_P)},
\end{equation}
where $e$ is the orbit's eccentricity, $L$ is the previously mentioned {\it semilatus rectum},
\begin{equation}
\frac{1}{L}=\frac{1}{2} \left( \frac{1}{r_+} + \frac{1}{r_-}\right),
\end{equation}
and $r_+$ and $r_-$ are the apoapsis and periapsis, {\it i.e.} the distances to the central body at aphelion and perihelion, respectively. The following relations are also valid,
\begin{eqnarray}\label{ellipseconditions}
\frac{d \phi}{dt} &=& \frac{\sqrt{GM_S L}}{r^2 }, \\ \nonumber \frac{dr}{dt} &=& e \sqrt{\frac{GM_S}{L}}\sin (\phi-\phi_P),\\ \nonumber \mathbf{r} \cdot \mathbf{v} &=& r \frac{dr}{dt} = \frac{e|\mathbf{h}| \sin (\phi-\phi_P)}{1 + e \cos (\phi-\phi_P)}, \\ \nonumber v^2 &=& \frac{GM_S}{L} \left[ 1 + e^2 + 2e \cos (\phi-\phi_P) \right] , \\ \nonumber |\mathbf{h}| &=& \sqrt{GM_S L}~~,~~|\mathbf{A}| = e GM_S.
\end{eqnarray}
The constants of motion of closed Newtonian orbits are not only the total energy and angular momentum (per mass), $\mathbf{h} = \mathbf{r} \times \mathbf{v}$, but also the Runge-Lenz vector,
\begin{equation}
\mathbf{A} = -\frac{GM_S \mathbf{r}}{r} + \mathbf{v} \times \mathbf{h},
\end{equation}
which points towards the perihelion.

Thus, in order to compute the precession of the latter due to a  small perturbing force, it suffices to obtain the (small) variation of the Runge-Lenz vector along the line perpendicular to both {\bf A} and the angular momentum,
\begin{equation}\label{variationphi0}
\frac{d \phi_P}{dt} = (\mathbf{h}\times \mathbf{A}) \cdot \frac{\frac{d\mathbf{A}}{dt}}{|\mathbf{h}| \mathbf{A}^2},
\end{equation}
using
\begin{equation}\label{variationRL}
\frac{d\mathbf{A}}{dt} = \mathbf{\Delta} \times \mathbf{h} + \mathbf{v} \times (\mathbf{r} \times \mathbf{\Delta}).
\end{equation}
Integrating, we can finally get
\begin{eqnarray}\label{precession}
\delta \phi_P &=& \int_0^{2\pi} \frac{d\phi_P }{dt} \frac{dt}{d\phi} d\phi = \\ \nonumber && \int_0^{2\pi} \frac{d\phi_P }{dt} \frac{L^2}{|\mathbf{h}| \left[ 1 + e \cos(\phi-\phi_P) \right]^2 } d\phi.
\end{eqnarray}

In the case under scrutiny, inserting Eq. (\ref{perturbativeforce}) into Eq. (\ref{variationRL}) yields
\begin{equation}
\frac{d\mathbf{A}}{dt} = \frac{\Delta_r}{r} \mathbf{r} \times \mathbf{h} + \frac{2\Delta_v}{v} \mathbf{v}\times \mathbf{h} ,
\end{equation}
so that Eq. (\ref{variationphi0}) becomes
\begin{eqnarray}
\nonumber && \frac{d \phi_P}{dt} = \frac{1}{e^2}\sqrt{\frac{L}{GM_S}}\left[ \Delta_r \left(1 - \frac{L}{r} \right) + 2\Delta_v \frac{ \mathbf{r} \cdot \mathbf{v} }{ rv } \right] = \\ && \frac{1}{e}\sqrt{\frac{L}{GM_S}} \times \\ \nonumber && \left[ \frac{ 2\Delta_v \sin (\phi-\phi_P) }{ \sqrt{ 1 + e^2 + 2e \cos (\phi-\phi_P) } } -\Delta_r \cos (\phi-\phi_P) \right] .
\end{eqnarray}
Notice that $\Delta_v \equiv -4 U' \dot{r} v / c^2 $ has no dependence on the additional parameters of the model under scrutiny.

In the following we consider the regime $\lambda\gg L \sim r$ and we Taylor expand the involved quantities
to second order in $r/\lambda$ (except in the nonrelativistic terms);
as such, using Eqs. (\ref{perturbativeforce}) and (\ref{precession}), we may write
\begin{equation}\label{precession0} \delta\phi_P =  \frac{6\pi GM_S }{ L c^2} + \frac{1-\theta}{3e} \int_0^{2\pi} I(\phi) \cos (\phi-\phi_P) d\phi, \end{equation}
\noindent with
\begin{widetext}
\begin{eqnarray}\label{integrand}
&& I(\phi) = (1-\theta) \left[ 1 + \frac{1}{10} \left( \frac{R_S}{\lambda}\right)^2 \right] \left( 1 + \frac{r}{\lambda} \right) \exp \left(-\frac{r}{\lambda}\right) + \\ \nonumber && \frac{GM_S}{R_S c^2} \Bigg\{ \theta \left[ 3\theta \left( 1 - \frac{\mu}{2}\right) + 2\nu \right] \left[ \frac{1}{2} \left( \frac{r}{R_S}\right)^2 - \left(\frac{\lambda}{R_S}\right)^2 \right]  + \\ \nonumber && \left( \frac{R_S}{\lambda}\right)^2 \Bigg[ \frac{2}{75} (1-\theta) \theta \left( \frac{R_S}{r}\right)^3 + \frac{1}{15}(1-\theta) [ \theta(\theta+3\mu-4) -2\nu-9 ] \frac{R_S}{r}  + \frac{1}{10}(1-e^2)(1-\theta)\frac{R_S}{L} - \\ \nonumber &&  \frac{3142 + 2465\mu}{2800} \theta^2 + \frac{3348+2400\mu + 865 \nu}{2100} \theta - \frac{276 + 220\nu + 135\mu}{525} + \\ \nonumber && (1-\theta) \left( \frac{10}{3} + \frac{\mu}{2} (1-\theta) - \theta \right) \frac{r}{R_S}  - \frac{1}{2} (1-e^2)(1-\theta) \frac{r}{L} \frac{r}{R_S} + \\ \nonumber && \frac{\theta^2(86+63\mu) - 4\theta ( 16+18\mu+9\nu) + 16(2+3\nu)}{120} \left( \frac{r}{R_S}\right)^2  + \frac{1}{8} \theta \left[ 3\theta \left(1 - \frac{\mu}{2}\right) + 2\nu \right] \left(\frac{r}{R_S}\right)^4 \Bigg] - \\ \nonumber && \frac{R_S}{3\lambda} \left[ (1-\theta) (\theta^2 + 6\theta \mu - 4\nu) + \theta \left[ 3\theta \left( 1 - \frac{\mu}{2}\right) + 2\nu \right] \left( \frac{r}{R_S}\right)^3 \right] + \\ \nonumber && (1-e^2) (1-\theta) \frac{R_S}{L} - \frac{\theta^2(86+63\mu) - 4 \theta(16+18\mu+9\nu) + 16(2+3\nu)}{60} + \\ \nonumber && (1-\theta)\left[ \theta(\theta + 3 \mu - 4 ) - 2 (9+\nu) \right] \frac{R_S}{3r} + \frac{4}{15}\theta(1-\theta) \left(\frac{R_S}{r}\right)^3 \Bigg\} , \nonumber
\end{eqnarray}
\end{widetext}
\noindent so that the familiar result from GR is recovered by setting $\theta = 1 $, as expected (except in the case of a perfectly circular orbit, $e=0$, when the perihelion is ill-defined).

In the above, the exponential contribution may be first expanded to third order in the eccentricity $e$
\begin{eqnarray} && \left( 1 + \frac{r}{\lambda} \right) \exp \left(-\frac{r}{\lambda}\right) \approx \exp \left(-\frac{L}{\lambda}\right)  \times \\ \nonumber && \Bigg[ 1 + \frac{L}{\lambda} + e \left(\frac{L}{\lambda}\right)^2 \cos (\phi-\phi_P) + \\ \nonumber && \frac{e^2}{2} \left(\frac{L}{\lambda}\right)^2 \left( \frac{L}{\lambda} - 3 \right) \cos^2 (\phi-\phi_P) + \\ \nonumber && \frac{e^3}{6} \left(\frac{L}{\lambda}\right)^2 \left( \left[\frac{L}{\lambda}\right]^2 - 8\frac{L}{\lambda} + 12 \right) \cos^3 (\phi-\phi_P) \Bigg], \end{eqnarray}
\noindent so that the third power leads to a contribution of second order in $e$ to Eq. (\ref{precession0}). The remaining terms in Eq. (\ref{integrand}) may be directly integrated, using
\begin{eqnarray} && \int_0^{2\pi} \frac{\cos x}{(1 + e \cos x)^n} dx = \\ \nonumber && \begin{cases} 3 \pi e \left( 1 + \frac{1}{4}e^2 \right)  & n = -3 \\ \pi e & n = -1 \\ 0 & n = 0 \\ \frac{2 \pi }{ e} \left( 1 -\frac{1 }{ \sqrt{1-e^2} } \right) \approx -\pi e \left( 1 + \frac{3}{4}e^2 \right) & n = 1 \\ -\frac{2 \pi e }{ (1-e^2)^{3/2}} \approx -2 \pi e \left( 1 + \frac{3}{2} e^2 \right) & n = 2 \\  -\frac{3 \pi e }{ (1-e^2)^{5/2}} \approx -3\pi e \left( 1 + \frac{5}{2} e^2 \right) & n = 3 \\ -\frac{ \pi e ( 4 + e^2) }{ (1-e^2)^{7/2}} \approx -4 \pi e \left( 1 + \frac{15}{4}e^2 \right) & n = 4 \end{cases} .
\end{eqnarray}
\noindent However, the ensuing expressions are too cumbersome, so we choose to instead also expand the ensuing integral to second order in $e$: the overall result is then given by
\begin{widetext}
\begin{eqnarray}
\label{deltaphi} \delta && \phi_P = \frac{6\pi GM_S }{ L c^2} + (1-\theta)^2\frac{\pi}{3}\left\{ 1 + e^2 \left[ \frac{3}{2} - \frac{L}{\lambda} + \frac{1}{8} \left( \frac{L}{\lambda}\right)^2 \right] \right\} \left[ 1 + \frac{1}{10} \left(\frac{R_S}{\lambda}\right)^2 \right] \left( \frac{L}{\lambda} \right)^2 \exp \left( -\frac{L}{\lambda}\right) + \\ \nonumber && (1-\theta) \frac{\pi GM_S}{12 L c^2} \Bigg\{ \bigg[ \theta \left[3\theta \left( 1 - \frac{\mu}{2} \right) + 2 \nu \right] \left[ -2(2 + 3e^2) + \left( 4 + 10 e^2 \right) \frac{L}{\lambda} - \left(2 + \frac{15}{2} e^2\right) \left(\frac{L}{\lambda} \right)^2 \right] - \\ \nonumber && \frac{1}{30}(2+3e^2) ( \theta^2 [86 + 63 \mu] -4 \theta [16 + 18\mu + 9\nu] + 16[2 + 3 \nu] ) \left( \frac{R_S}{\lambda}\right)^2 \bigg] \left( \frac{L}{R_S}\right)^3 - \\ \nonumber && (1-\theta)\left( \frac{28}{3} + 2 ( 1 - \theta) \mu -4 \theta + e^2 \left[ 8 + \frac{3}{2} \left( \mu[1 - \theta] - 2\theta \right)  \right] \right) \left( \frac{L}{\lambda}\right)^2 + \\ \nonumber && \frac{4}{5} (1-\theta) \theta (4+e^2) \left[ 1 + \frac{1}{10}\left( \frac{R_S}{\lambda}\right)^2 \right] \left(\frac{R_S}{L} \right)^2 + \frac{4}{3} (1 - \theta) \left( \left[ \theta(\theta-4+3\mu) -2\nu-9 \right] \left[ 1 + \frac{1}{5} \left( \frac{R_S}{\lambda} \right)^2 \right] -9 \right) \Bigg\}.
\end{eqnarray}
\end{widetext}
Notice that the above collapses to $\delta \phi_P = 4\pi GM_S/Lc^2$ when the model parameters $\theta$, $\mu$ and $\nu$ vanish and $\lambda \to \infty$: this falls short of the GR prediction of $\delta \phi_P = 6\pi GM_S/Lc^2$ by a factor $2/3$.

The prediction for the precession of the perihelion assuming a PPN metric \cite{Wi} together with the Newtonian effect of a quadrupole moment $J_2 \sim (2.2 \pm 0.1 ) \times 10^{-7}$ \cite{J2} of the Sun is given by
\begin{equation} \delta \phi_P = \left[ \frac{2(1+\gamma) - \beta }{3} + 3 \times 10^3 J_2 \right] \frac{6\pi GM_S}{Lc^2}, \label{deltaphiPPN} \end{equation}
\noindent with the most stringent bounds on the PPN parameters $\beta $ \cite{Messenger} and $\gamma$ \cite{Cassini} given by
\begin{eqnarray} \label{PPNconstraints} \beta-1 &=& (-4.1\pm 7.8) \times 10^{-5} , \\ \nonumber \gamma - 1 &=& (2.1 \pm 2.3) \times 10^{-5}.\end{eqnarray}
The bound on $\beta$ results from recent observations of Mercury, 
including data from the Messenger spacecraft.

The result $\delta \phi_P = 4\pi GM_S/Lc^2$ is equivalent to having $\beta = 2\gamma$ \cite{Wi}. In particular, this is precisely what stems from the extraneous comparison of $f(R)$ models with a Brans-Dicke theory with parameter $\omega = 0$, which incorrectly leads to $\gamma = 1/2$ and $\beta=1$.

Conversely, inspection shows that setting $\theta=1$ immediately yields the GR prediction for the precession of the perihelion, independently of the remaining model parameters: this reflects the dependence of the model parameters $\alpha_0, \beta_i, \zeta_i \sim 1-\theta$, and confirms the previous findings of Ref. \cite{CPM} --- where it was noted that the vanishing of the zeroth-order coupling $\alpha_0=0$ when $\theta=1$ evades the stringent constraints of Yukawa forces existing for characteristic lengthscales $ 1~{\rm mm} < \lambda < 1000~{\rm AU} $ \cite{Fischbach}.

\begin{figure}[ht]
\centering
\includegraphics[width=0.5\textwidth]{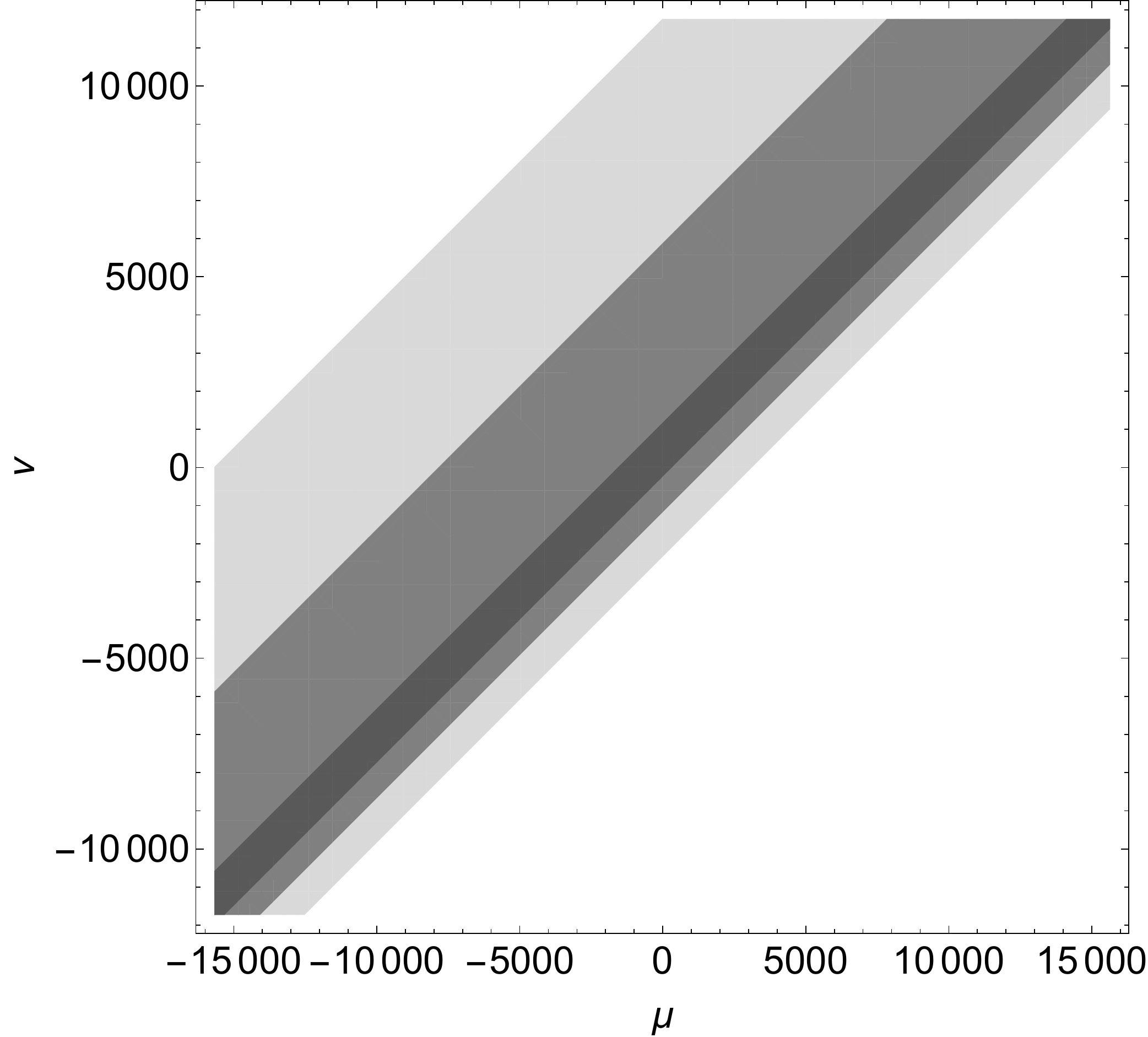}
\caption{Exclusion plot for the model parameters $(\mu,\nu)$, for $\lambda = 50L$ and $\theta = \{1 + 10^{-13}, 1 + 2 \times 10^{-13},1 + 10^{-12} \}$ (light, medium, dark grey).}
\label{plot1}
\end{figure}

\begin{figure}[ht]
\centering
\includegraphics[width=0.5\textwidth]{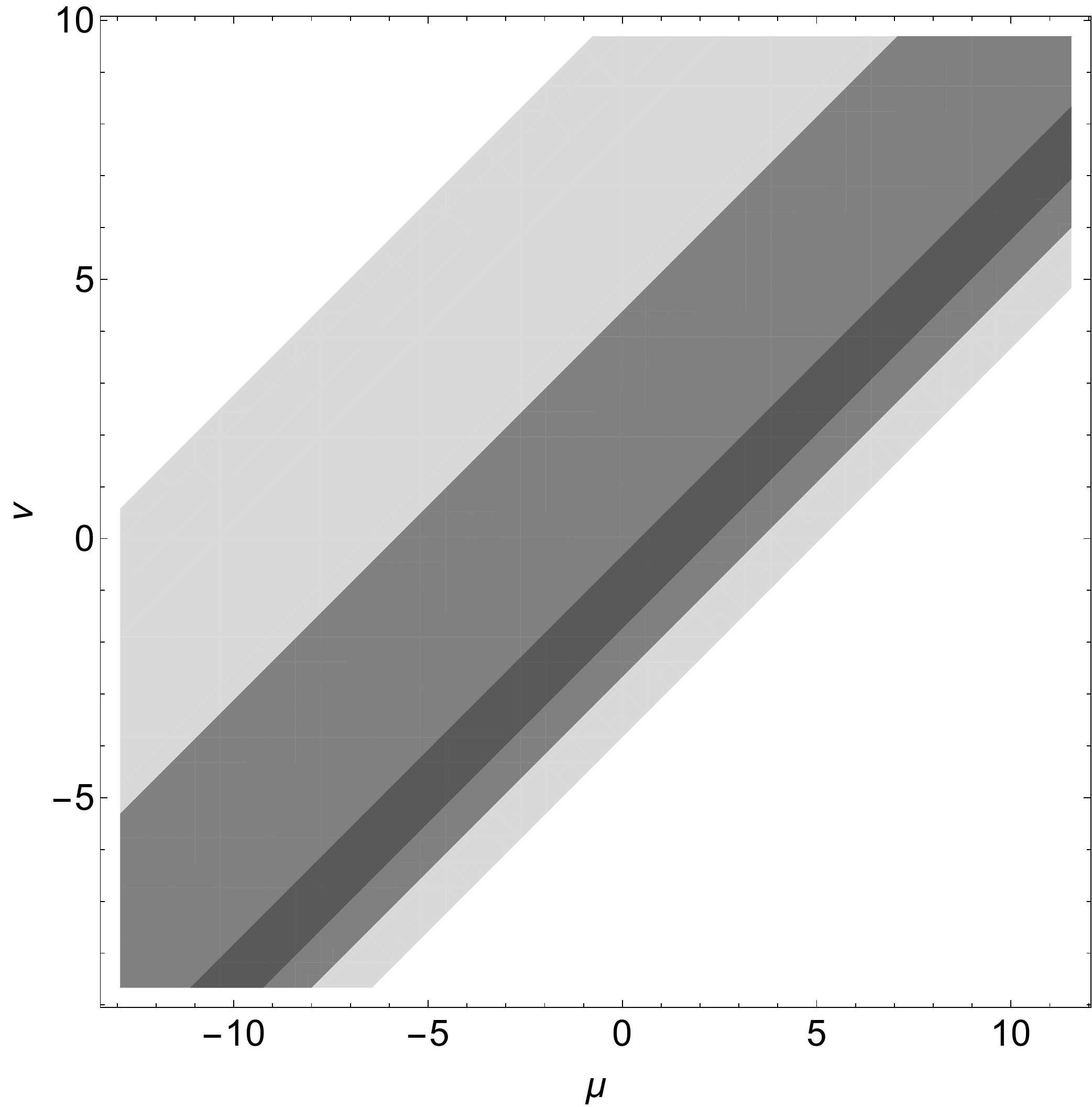}
\caption{Exclusion plot for the model parameters $(\mu,\nu)$, for $\lambda = 50L$ and $\theta = \{1 + 10^{-10}, 1 + 2 \times 10^{-10},1 + 10^{-9} \}$ (light, medium, dark grey).}
\label{plot2}
\end{figure}

\begin{figure}[ht]
\centering
\includegraphics[width=0.5\textwidth]{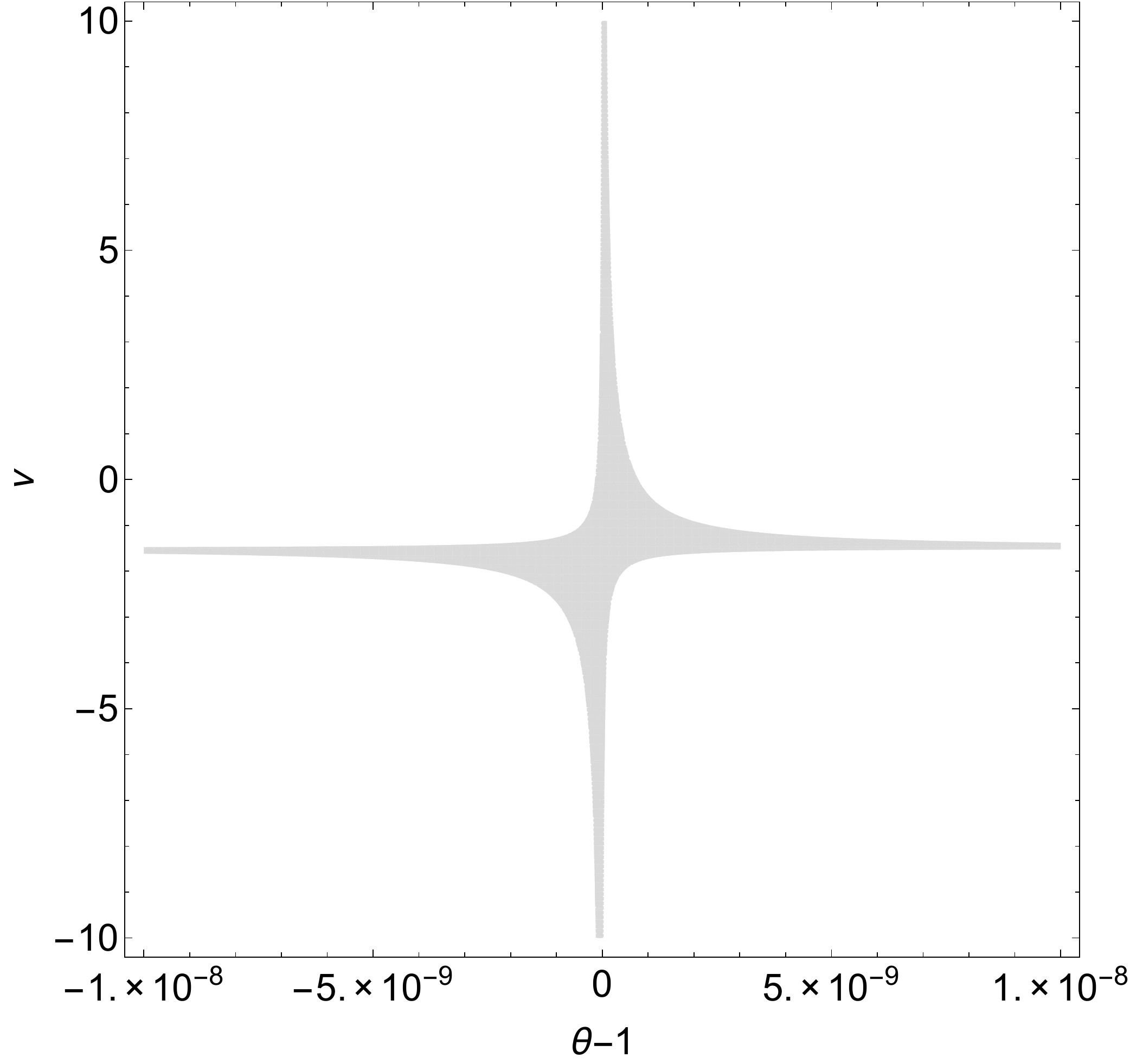}
\caption{Exclusion plot for the model parameters $(\theta,\nu)$, for $\lambda = 50L$ and $\mu = 0 $.}
\label{plot3}
\end{figure}

\begin{figure}[ht]
\centering
\includegraphics[width=0.5\textwidth]{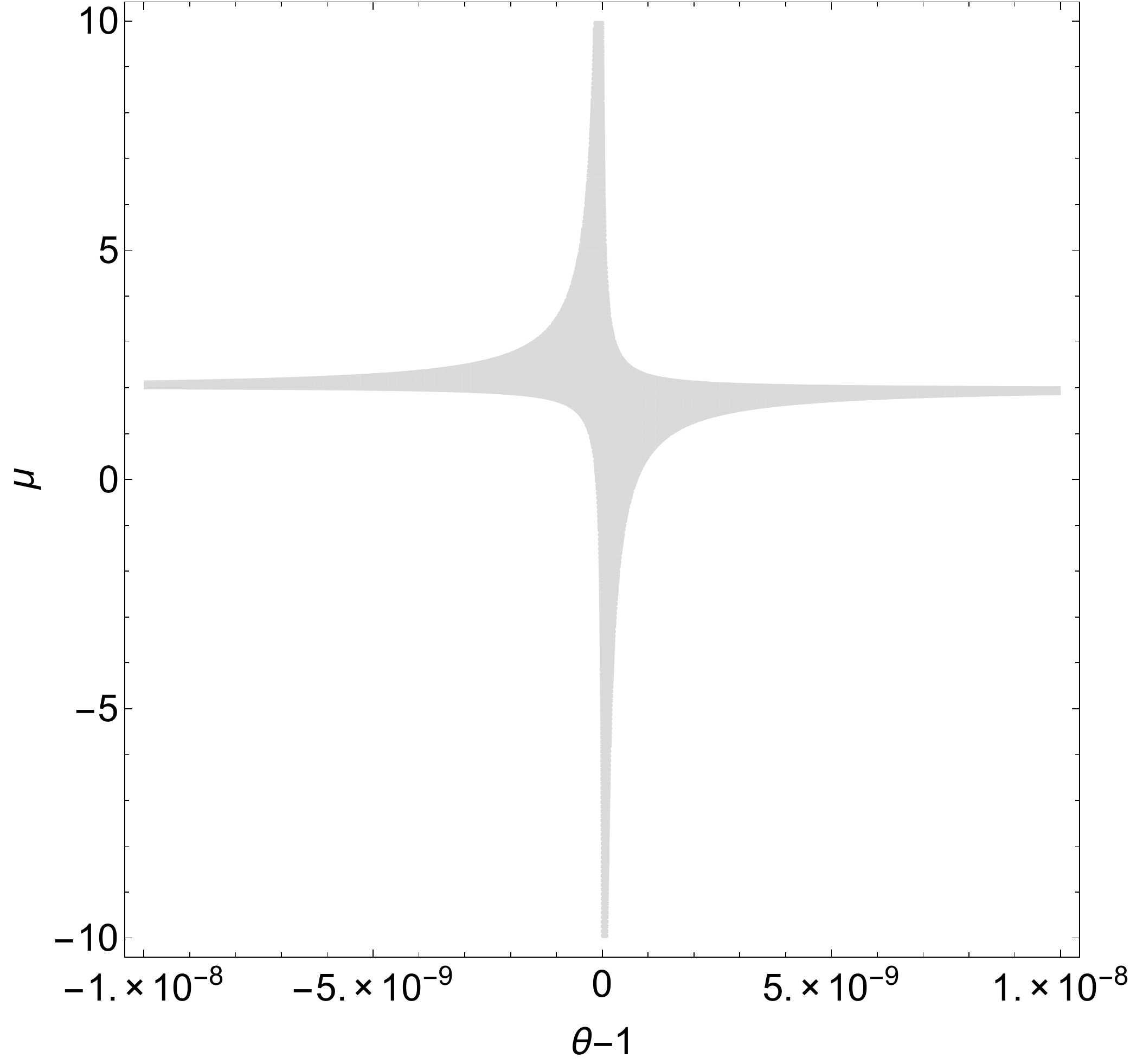}
\caption{Exclusion plot for the model parameters $(\theta,\mu)$, for $\lambda = 50L$ and $\nu = 0 $.}
\label{plot4}
\end{figure}

\begin{figure}[ht]
\centering
\includegraphics[width=0.5\textwidth]{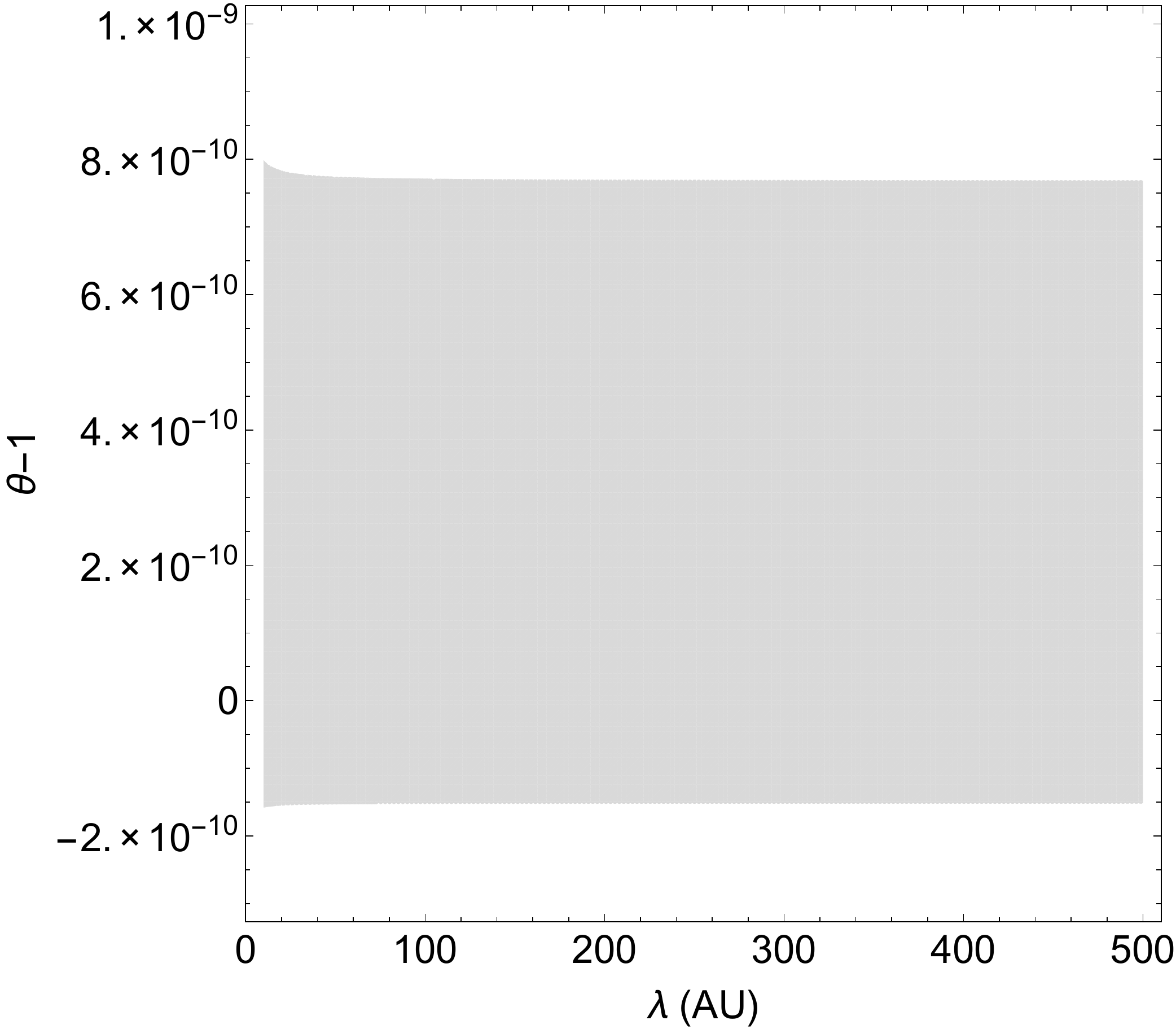}
\caption{Exclusion plot for the model parameters $(\lambda,\theta)$, for $\mu = \nu = 0 $.}
\label{plot5}
\end{figure}

\begin{figure}[ht]
\centering
\includegraphics[width=0.5\textwidth]{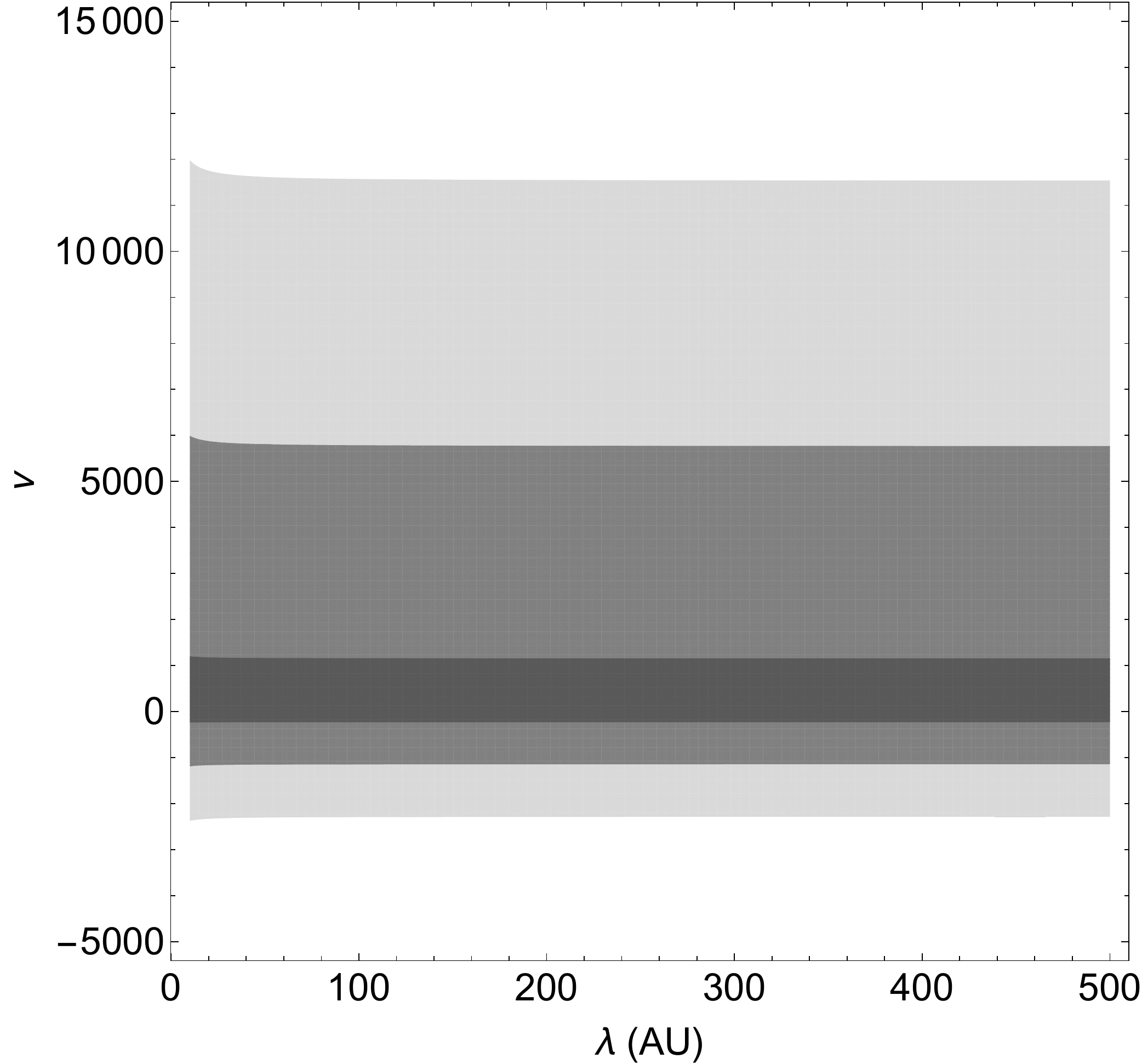}
\caption{Exclusion plot for the model parameters $(\lambda,\nu)$, for $\mu = 0 $ and $\theta = \{1 + 10^{-13}, 1 + 2 \times 10^{-13},1 + 10^{-12} \}$ (light, medium, dark grey).}
\label{plot6}
\end{figure}

\begin{figure}[ht]
\centering
\includegraphics[width=0.5\textwidth]{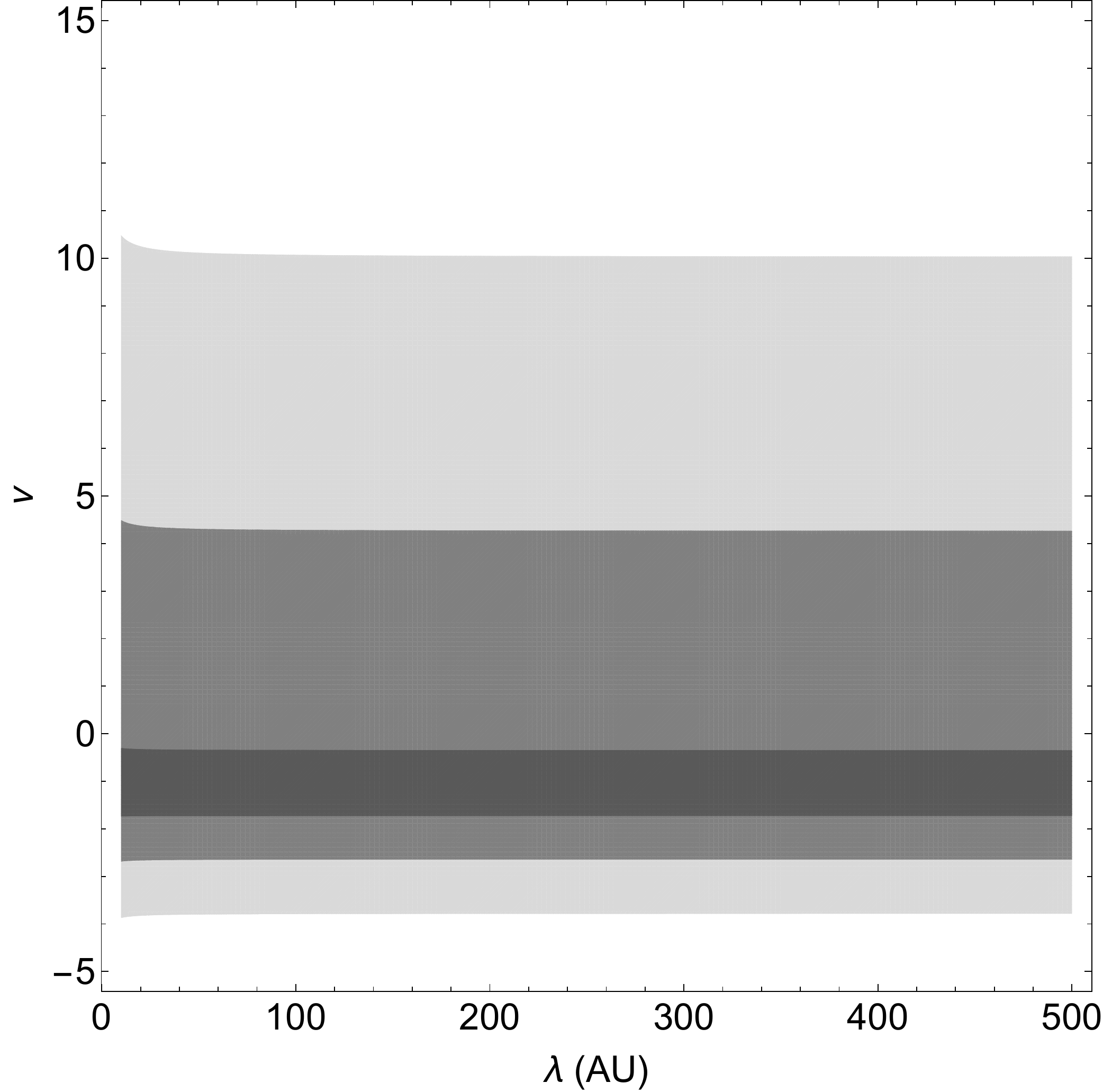}
\caption{Exclusion plot for the model parameters $(\lambda,\nu)$, for $\mu = 0 $ and $\theta = \{1 + 10^{-10}, 1 + 2 \times 10^{-10},1 + 10^{-9} \}$ (light, medium, dark grey).}
\label{plot7}
\end{figure}

Inserting the values for the mass of the Sun, $M_\odot = 1.989\times 10^{30}$ kg and the {\it semilatus rectum} of Mercury, $L = 5.546 \times 10^7~{\rm m}$, together with the experimental bounds for the PPN parameters $\beta $ and $\gamma$ given in Eq. (\ref{PPNconstraints}), we find that the additional perihelion precession due to the model under scrutiny is bounded by
\begin{equation} - 5.87537 \times 10^{-4} < \delta \phi_P - 42.98'' < 2.96635 \times 10^{-3} , \end{equation}
\noindent so that Eq. (\ref{deltaphi}) for $\delta \phi_P$ allows us to obtain exclusion plots for the four independent quantities $\theta = q_1/a_2$, $\mu = a_3/a_2^2$, $\nu = q_2/a_2^2$ and $R_S/\lambda =R_S/\sqrt{6a_2} \ll 1$, as depicted on Figs. \ref{plot1}-\ref{plot7}, using the previously considered experimental bounds for $\beta $ and $\gamma$.

The BepiColombo mission offers the best short-term possibility for tightening current constraints on the PPN parameters, shown in Eq. (\ref{PPNconstraints}): indeed, the radioscience experiment onboard the spacecraft is expected to yield an  order of magnitude improvement on $\beta$ \cite{BepiColomboBeta} and $\gamma$ \cite{BepiColomboGamma},
\begin{eqnarray} \label{PPNconstraintsBepiColombo} |\beta-1| &\leq& 7.81 \times 10^{-6} , \\ \nonumber |\gamma - 1| &\leq& 5.07 \times 10^{-6}.\end{eqnarray}
Using this figures to derive the allowed range for the model parameters mentioned above does not change the corresponding exclusion plots qualitatively, but naturally leads to a reduction on their admissible bounds of approximately one order of magnitude.
\section{Conclusions}
In this work we have computed the metric solutions for a NMC gravity model around a Minkowski background. It is shown that, up to order $O(1/c^4)$, the corrections depend on the $f^1(R)$ and $f^2(R)$ functions and cannot be expressed in terms of powers of $1/r$: indeed, it is found that the obtained solutions must be expressed in the PPNY approximation, as first proposed in Ref. \cite{CPM}.

This opens up the possibility of addressing a wider class of physical situations with great accuracy. Furthermore, the results obtained in this work might be relevant for distinguishing between GR, $f(R)$ and non minimally coupled theories from the analysis of detailed observations data in the future.
\appendix
\section*{Appendix A}

In order to compute $h_{00}^{(4)}$ we need the corresponding term $R^{(4)}$ in the expansion Eq. (\ref{R-expansion}) of the
Ricci scalar; this can be obtained by solving the trace Eq. (\ref{trace}) at order
$O\left(1\slash c^4\right)$.

In the following, in order to avoid a cumbersome notation, we replace the symbol $R^{(2)}$ with $R$.
Using the gauge conditions Eq. (\ref{gauge-1}),
the trace of the field equations at order $O\left(1\slash c^4\right)$ yields the following equation
for $R^{(4)}$:
\begin{widetext}
\begin{eqnarray}\label{R4-equation}
& &\nabla^2 R^{(4)} - \frac{1}{6a_2} R^{(4)} - \frac{1}{c^2} R_{,00} + \frac{3a_3}{2a_2} \nabla^2 R^2 +
2a_2R\nabla^2 R - \frac{2}{c^2} U\nabla^2 R + \frac{64\pi G}{c^4}\left(a_2-q_1\right)\frac{q_1}{a_2} U_{,ij}\rho_{,ij}  \nonumber\\
& & - \frac{16\pi G}{c^2}\left[ q_1\left(\rho\nabla^2 R +
R\nabla^2\rho\right) + \frac{q_2}{a_2} \nabla^2(\rho R) \right]
+ 24a_2^2 R_{,ij}R_{,ij} - \frac{8}{c^2}\left(a_2-q_1\right)U_{,ij}R_{,ij} \nonumber\\
& & - \frac{384\pi G}{c^2} a_2q_1 \rho_{,ij} R_{,ij} - \frac{4\pi G}{c^2} \frac{q_1}{a_2} \rho R +
\frac{8\pi G}{c^4} \frac{q_1}{a_2} \rho_{,00} + \frac{16\pi G}{c^4} \frac{q_1}{a_2} U\nabla^2\rho \nonumber\\
& & +
\frac{128\pi^2 G^2}{c^4}\frac{q_1^2}{a_2} \rho\nabla^2\rho +
\frac{1536\pi^2 G^2}{c^4} q_1^2\rho_{,ij}\rho_{,ij} = - \frac{4\pi G}{3a_2c^4}\left( \rho\Pi - 3p \right).
\end{eqnarray}
\end{widetext}
Next, we rewrite this equation in the form of a Yukawa equation of the type
\clearpage
\begin{equation}\label{Yukawa}
\left( \nabla^2 - m^2 \right) \left( R^{(4)} + \dots \right) = -4\pi Q,
\end{equation}
where we recall that $m^2 = 1\slash 6a_2$ and we introduce the potential \cite{Cl}
\begin{equation}
X(Q) = \int Q(t,\y) \frac{e^{-m|\x-\y|}}{|\x-\y|} d^3y,
\end{equation}
which solves the equation
\begin{equation}
\left( \nabla^2 - m^2 \right) X(Q) = -4\pi Q.
\end{equation}
In order to put Eq. (\ref{R4-equation}) into the form (\ref{Yukawa}), we make use of the following identity
for two arbitrary potentials $\tilde U$ and $\tilde V$:
\begin{eqnarray}\label{identity}
\tilde U_{,ij}\tilde V_{,ij} = \frac{1}{2} & \bigg[ & \nabla^2(\nabla\tilde U \cdot \nabla\tilde V) \\ \nonumber && -
\nabla\tilde U \cdot \nabla (\nabla^2\tilde V) - \nabla\tilde V \cdot \nabla (\nabla^2\tilde U) \bigg].
\end{eqnarray}
Using this identity, the trace Eq. (\ref{trace-2}), and the Poisson equation for the Newtonian potential,
$\nabla^2U=-4\pi G\rho$, we get the following relations:
\begin{widetext}
\begin{eqnarray}\label{ij-identities}
U_{,ij}R_{,ij} &=& \frac{1}{2}\left( \nabla^2 - \frac{1}{6a_2} \right)\nabla U \cdot \nabla R +
2\pi G\nabla\rho \cdot \nabla R + \frac{2\pi G}{3a_2c^2} \nabla U \cdot \nabla \left(\rho-6q_1\nabla^2\rho\right),
\\ \nonumber
R_{,ij}R_{,ij} &=& \frac{1}{2} \left( \nabla^2 - \frac{1}{6a_2} \right)\left( |\nabla R|^2 -
\frac{R^2}{12a_2} \right) + \frac{R^2}{144a_2^2} + \frac{\pi G}{3a_2^2c^2} \left[ 2 q_1 R\nabla^2\rho - \frac{\rho R}{3} + 4 \nabla R \cdot \nabla \left(\rho-6q_1\nabla^2\rho\right) \right], \\ \nonumber
R_{,ij}\rho_{,ij} &=& \frac{1}{2}\left( \nabla^2 - \frac{1}{6a_2} \right)\nabla\rho \cdot \nabla R -
\frac{1}{2} \nabla R \cdot \nabla(\nabla^2\rho) +
\frac{2\pi G}{3a_2c^2} \nabla\rho \cdot \nabla \left(\rho-6q_1\nabla^2\rho\right), \nonumber\\
U_{,ij}\rho_{,ij} &=& \frac{1}{2}\left( \nabla^2 - \frac{1}{6a_2} \right)\nabla\rho \cdot \nabla U +
\frac{1}{12a_2} \nabla\rho \cdot \nabla U + 2\pi G|\nabla\rho|^2 -
\frac{1}{2} \nabla U \cdot \nabla(\nabla^2\rho), \nonumber\\
\rho_{,ij}\rho_{,ij} &=& \frac{1}{2}\left( \nabla^2 - \frac{1}{6a_2} \right)|\nabla\rho|^2 +
\frac{1}{12a_2}|\nabla\rho|^2 - \nabla\rho \cdot \nabla(\nabla^2\rho). \nonumber
\end{eqnarray}
\end{widetext}
Now we introduce the potential \cite{Cl}
\begin{equation}
\hat\chi = G\int \rho(t,\y)e^{-m|\x-\y|}d^3y.
\end{equation}
Using the solution for the trace equation at second order, Eq. (\ref{R2-solution}) , one can show that the potential
$\hat\chi$ satisfies the equation
\begin{eqnarray}\label{chi-potential}
&& \left( \nabla^2 - \frac{1}{6a_2} \right)\hat\chi = \\ \nonumber && -c^2\sqrt{6a_2}
\left( 1 - \frac{q_1}{a_2} \right)^{-1}\left( R - \frac{8\pi G}{c^2} \frac{q_1}{a_2} \rho \right).
\end{eqnarray}
Then, using the relations (\ref{ij-identities}), the trace equation (\ref{trace-2}), and transforming
the quantities $R\nabla^2\rho$, $U\nabla^2\rho$ and $\rho\nabla^2\rho$ by means of the identity
$\nabla^2(ab)=a\nabla^2 b + b\nabla^2 a + 2\nabla a \cdot \nabla b$, we put Eq. (\ref{R4-equation})
for $R^{(4)}$ in the Yukawa form Eq. (\ref{Yukawa}). The solution of the resulting equation is
\begin{widetext}
\begin{eqnarray}\label{R4-solution}
R^{(4)} &=& - \frac{1}{c^4\sqrt{6a_2}}\left( 1 - \frac{q_1}{a_2} \right)\hat\chi_{,00} -
\left( \frac{3a_3}{2a_2} - a_2 \right)R^2 - \frac{16\pi G}{c^2}\left( q_1 - \frac{q_2}{a_2} \right)\rho R \\
&+&  \frac{64\pi^2G^2}{c^4} \frac{q_1^2}{a_2}\rho^2
- 12a_2^2|\nabla R|^2 + \frac{4}{c^2}\left(a_2 - q_1\right)\nabla U \cdot \nabla R
+ \frac{192\pi G}{c^2} a_2q_1\nabla\rho \cdot \nabla R \nonumber\\
&-& \frac{32\pi G}{c^4} \frac{q_1}{a_2}\left(a_2 - q_1\right)\nabla\rho \cdot \nabla U
-\frac{768\pi^2G^2}{c^4}q_1^2|\nabla\rho|^2 - \frac{1}{12\pi a_2c^2} X(UR) \nonumber\\
&+& \frac{1}{8\pi}\left( \frac{a_3}{2a_2^2}+1 \right)X(R^2) + \frac{2G}{3a_2c^4} X(\rho U) -
\frac{G}{12c^2}\left( 16+20\frac{q_1}{a_2}+8\frac{q_2}{a_2^2} \right)X(\rho R)\nonumber\\
&+& \frac{8\pi G^2}{3c^4} \frac{q_1}{a_2}\left( 4-\frac{q_1}{a_2} \right)X(\rho^2)
- \frac{4G}{3c^4}\left[ 1-\frac{q_1}{a_2}\left(2-\frac{q_1}{a_2}\right)\right]X(\nabla\rho \cdot \nabla U)
\nonumber\\ \nonumber
&+& \frac{4G}{c^2}\left(a_2-q_1\right)X(\nabla\rho \cdot \nabla R) -
\frac{32\pi G^2}{c^4} q_1\left(1-\frac{q_1}{a_2}\right)X(|\nabla\rho|^2) - \frac{G}{a_2c^4} X(p) + \frac{G}{3a_2c^4} X(\rho\Pi).
\end{eqnarray}
\end{widetext}
We can now write the $0-0$ component of the field Eqs. (\ref{field-eqs})
at order $O\left(1\slash c^4\right)$. Using the expressions for $R_{00}$ and $T_{00}$ given by Eqs.
(\ref{R-00}) and $(\ref{T-00})$, respectively, and Eq. (\ref{R4-equation}) to eliminate the term
proportional to $R^{(4)}$, we find that $h_{00}^{(4)}$ obeys the following:
\begin{widetext}
\begin{eqnarray}\label{h00-4-equation}
& & -\frac{1}{2} \nabla^2 h_{00}^{(4)} - \frac{1}{c^4}\nabla^2 U^2 + \left(\frac{3}{2}a_3-2a_2^2\right)\nabla^2 R^2
- 3\frac{a_2}{c^2} \nabla^2(UR) + \frac{24\pi G}{c^4} q_1 \nabla^2(\rho U) \nonumber\\
& & + \frac{16\pi G}{c^2}\left(2a_2q_1-q_2\right)\nabla^2(\rho R) -
\frac{128\pi^2G^2}{c^4} q_1^2\nabla^2\rho^2 - 18 \frac{a_2^2}{c^2}\nabla^2 R_{,00} +
\frac{6}{c^4}\left(a_2-q_1\right)\nabla^2 U_{,00} \nonumber\\
& & + \frac{144\pi G}{c^4} a_2q_1\nabla^2\rho_{,00} + a_2\nabla^2 R^{(4)} +
\frac{8}{c^4}\left(a_2-q_1\right)U_{,ij}U_{,ij} - 24\frac{a_2^2}{c^2}U_{,ij}R_{,ij} \nonumber\\
& & + \frac{192\pi G}{c^4} a_2q_1 U_{,ij}\rho_{,ij} + \frac{1}{6c^2} UR + \frac{2}{3} a_2R^2 -
\frac{28\pi G}{3c^4} \rho U + \frac{20\pi G}{c^2}\left(\frac{a_2}{3}-q_1\right)\rho R  - \frac{96\pi^2G^2}{c^4} q_1\rho^2
\nonumber\\ & &  =   \frac{4\pi G}{c^4} \left[ \rho(\Pi + 2 v^2) + 3 p \right] .
\end{eqnarray}
\end{widetext}
This can be written in the form of a Poisson equation of the type
\begin{equation}\label{Poisson}
\nabla^2 \left( h_{00}^{(4)} + \dots \right) = -4\pi Q.
\end{equation}
Moreover, we denote by $\V$ the Poisson integral:
\begin{equation}
\V(Q) = \int \frac{Q(t,\y)}{|\x-\y|} d^3y ,
\end{equation}
so that $\nabla^2 \V(Q) = -4\pi Q$. We proceed as in the computation of $R^{(4)}$:
using the identity (\ref{identity}), the trace Eq. (\ref{trace-2}), and the Poisson equation for
the Newtonian potential, $\nabla^2U=-4\pi G\rho$, we get the following relations,
\begin{widetext}
\begin{eqnarray}\label{ij-identities-bis}
U_{,ij}U_{,ij} &=& \frac{1}{2} \nabla^2(|\nabla U|^2) + 4\pi G \nabla\rho \cdot \nabla U, \\ \nonumber
U_{,ij}R_{,ij} &=& \frac{1}{2} \nabla^2(\nabla U \cdot \nabla R) - \frac{1}{24a_2} \nabla^2(UR)+  2\pi G \nabla\rho \cdot \nabla R +
\frac{\pi G}{3c^2} \frac{q_1}{a_2^2} \nabla^2(\rho U) + \frac{2\pi G}{3a_2c^2}\left(1-\frac{q_1}{a_2}\right)
\nabla\rho \cdot \nabla U \\ \nonumber
&-& \frac{4\pi G}{c^2} \frac{q_1}{a_2} \nabla U \cdot \nabla(\nabla^2\rho)  - \frac{\pi G}{6a_2} \rho R + \frac{1}{144a_2^2} UR - \frac{\pi G}{18a_2^2c^2} \rho U + \frac{4\pi^2G^2}{3c^2} \frac{q_1}{a_2^2}\rho^2, \nonumber \\ \nonumber
U_{,ij}\rho_{,ij} &=& \frac{1}{2} \nabla^2(\nabla\rho \cdot \nabla U) + 2\pi G|\nabla\rho|^2 - \frac{1}{2} \nabla U \cdot \nabla(\nabla^2\rho). \nonumber
\end{eqnarray}
\end{widetext}
Then, using relations (\ref{ij-identities-bis}), Eq. (\ref{h00-4-equation})
for $h_{00}^{(4)}$ can be recast in the Poisson form Eq. (\ref{Poisson}), with solution
\begin{widetext}
\begin{eqnarray}
h_{00}^{(4)} &=& -\frac{2}{c^4} U^2 + \left(3a_3-4a_2^2\right)R^2 -4\frac{a_2}{c^2} UR +
\frac{32\pi G}{c^4} q_1\rho U + \frac{32\pi G}{c^2}\left(2a_2q_1-q_2\right)\rho R \\ \nonumber
&-& \frac{256\pi^2G^2}{c^4} q_1^2\rho^2 - 36\frac{a_2^2}{c^2} R_{,00} + \frac{12}{c^4}\left(a_2-q_1\right)U_{,00}
+\frac{288\pi G}{c^4} a_2q_1\rho_{,00} \\ \nonumber
&+& \frac{8}{c^4}\left(a_2-q_1\right)|\nabla U|^2
- 24\frac{a_2^2}{c^2}\nabla U \cdot \nabla R + \frac{192\pi G}{c^4} a_2q_1 \nabla\rho \cdot \nabla U
-\frac{a_2}{3\pi} \V(R^2) + \frac{4G}{c^4} \V(\rho U) \\ \nonumber
&-& \frac{2G}{c^2}\left(\frac{8}{3} a_2-5q_1\right)\V(\rho R)
+ \frac{64\pi G^2}{c^4} q_1\V(\rho^2) - \frac{8G}{c^4}\left(a_2-q_1\right)\V(\nabla\rho \cdot \nabla U) \\ \nonumber
&+& 24\frac{G}{c^2} a_2^2 \V(\nabla\rho \cdot \nabla R) - \frac{192\pi G^2}{c^4} a_2q_1 \V(|\nabla\rho|^2)
+ \frac{2G}{c^4} \V(\rho\Pi) + \frac{4G}{c^4} \V(\rho v^2) + \frac{6G}{c^4} \V(p) + 2a_2 R^{(4)}.
\end{eqnarray}
\end{widetext}
Substituting in the above expression Eq. (\ref{R4-solution}) for $R^{(4)}$, we finally obtain the solution
for $h_{00}^{(4)}$ given in Eq. (\ref{h004}) of Section \ref{subsec:h004}.

\appendix
\section*{Appendix B}

The component $g_{00}$ of the metric in Eq. (\ref{g00}) contains contributions with the potentials $\V$ and $X$
depending on the gradient terms $\nabla\rho \cdot \nabla U$, $\nabla\rho \cdot \nabla R$ and $|\nabla\rho|^2$.
Following Ref. \cite{Cl}, we transform such potentials into expressions without gradient terms.
Arguing as in Section \ref{sec:gauge}, we find the following identities:
\begin{eqnarray}
&& G\V(\nabla\rho \cdot \nabla U) = 4\pi G^2 \V(\rho^2) + \psi_1, \\ \nonumber
&& G\V(\nabla\rho \cdot \nabla R) = \\ \nonumber &&  \frac{8\pi G^2}{c^2} \frac{q_1}{a_2} \V(|\nabla\rho|^2) +
\frac{4\pi G^2}{3c^2} \frac{a_2-q_1}{a_2^2} \V(\rho^2)  \\ \nonumber && - \frac{\left(a_2-q_1\right)}{3c^2a_2^2}\left( \frac{1}{6a_2} \psi_0 - \frac{1}{\sqrt{6a_2}} \psi_2 - \psi_3 \right), \\ \nonumber
&& G X(\nabla\rho \cdot \nabla U) = 4\pi G^2 X(\rho^2) + \frac{1}{\sqrt{6a_2}} \psi_4 + \psi_5, \\ \nonumber
&& G X(\nabla\rho \cdot \nabla R) = \frac{8\pi G^2}{c^2} \frac{q_1}{a_2} X(|\nabla\rho|^2)  \\ \nonumber && +
\frac{4\pi G^2}{3c^2} \frac{a_2-q_1}{a_2^2} X(\rho^2)  \\ \nonumber && + \frac{\left(a_2-q_1\right)}{3c^2a_2^2}\left[ \psi_6 + \frac{\psi_7+\psi_8}{\sqrt{6a_2}} +
\frac{\psi_9-\psi_{10}}{6a_2} \right].
\end{eqnarray}
Substituting these identities into Eq. (\ref{g00}) for $g_{00}$, the terms proportional to
$\V(|\nabla\rho|^2)$ and $X(|\nabla\rho|^2)$ cancel exactly. The eleven potentials $\psi_0,\dots,\psi_{10}$
appearing in the previous identities are given by
\begin{equation}
\psi_i(t,\x) = G^2\int \frac{\rho(t,\y)\rho(t,\z)}{|\x-\y| |\y-\z|} \Psi_i (\x,\y,\z)d^3yd^3z, \\
\end{equation}
with
\begin{eqnarray}
\Psi_0(\x,\y,\z) &=& e^{-m|\y-\z|} , \\ \nonumber
\Psi_1(\x,\y,\z) &=& \frac{(\x-\y)\cdot(\y-\z)}{|\x-\y|^2|\y-\z|^2} , \\ \nonumber
\Psi_2(\x,\y,\z) &=&  \frac{(\x-\y)\cdot(\y-\z)}{|\x-\y|^2|\y-\z|}
e^{-m|\y-\z|}, \\ \nonumber
\Psi_3 (\x,\y,\z)&=&  \frac{(\x-\y)\cdot(\y-\z)}{|\x-\y|^2|\y-\z|^2}
e^{-m|\y-\z|}, \\ \nonumber
\Psi_4(\x,\y,\z) &=&  \frac{(\x-\y)\cdot(\y-\z)}{|\x-\y||\y-\z|^2}
e^{-m|\x-\y|}, \\ \nonumber
\Psi_5(\x,\y,\z) &=&  \frac{(\x-\y)\cdot(\y-\z)}{|\x-\y|^2|\y-\z|^2}
e^{-m|\x-\y|}, \\ \nonumber
\Psi_6 (\x,\y,\z)&=&  \frac{(\x-\y)\cdot(\y-\z)}{|\x-\y|^2|\y-\z|^2}
e^{-m(|\x-\y|+|\y-\z|)}, \\ \nonumber
\Psi_7(\x,\y,\z) &=&  \frac{(\x-\y)\cdot(\y-\z)}{|\x-\y|^2|\y-\z|}
e^{-m(|\x-\y|+|\y-\z|)}, \\ \nonumber
\Psi_8 (\x,\y,\z)&=&  \frac{(\x-\y)\cdot(\y-\z)}{|\x-\y||\y-\z|^2}
e^{-m(|\x-\y|+|\y-\z|)}, \\ \nonumber
\Psi_9 (\x,\y,\z)&=&  \frac{(\x-\y)\cdot(\y-\z)}{|\x-\y||\y-\z|}
e^{-m(|\x-\y|+|\y-\z|)}, \\ \nonumber
\Psi_{10}(\x,\y,\z) &=& e^{-m(|\x-\y|+|\y-\z|)}.
\end{eqnarray}
The potentials $\psi_1,\dots,\psi_9$ coincide with those found in Ref. \cite{Cl}.

\section*{Appendix C}

We list the expressions of the potentials appearing in the $g_{00}$ coefficient of the PPNY metric,
evaluated for $r>R_S$ under the assumptions given in Section \ref{sec:spheric-metric} and
Subsection \ref{subsec:Yukawa}.
\begin{eqnarray}\label{Ypotentialhomogeneous}
\Y &=& GM_S \frac{e^{-r/\lambda}}{r} \left[ 1 + \frac{1}{10}\left(\frac{R_S}{\lambda}\right)^2 \right], \\ \nonumber
\Sigma_2 &=&  \frac{6}{5} \frac{(GM_S)^2}{R_S} \frac{e^{-r/\lambda}}{r} \left[ 1 + \frac{2}{21}\left(\frac{R_S}{\lambda}\right)^2 \right].
\end{eqnarray}
In the case of uniform density $\rho$ we have $\Pi=0$, hence $\Sigma_3=0$. At the required order
the pressure is given by Newtonian equilibrium:
$p(r)=p(0)(1-r^2/R_S^2)$ where the pressure $p(0)$ at the center of the body is
$p(0)=G(\pi/6)^{1/3}M_S^{2/3}\rho^{4/3}$. That yields for the potential $\Sigma_4$
\begin{eqnarray}
& & \Sigma_4 =
\frac{e^{-r/\lambda}}{r}\frac{1}{5}\frac{(GM_S)^2}{R_S}
\left[ 1 + \frac{1}{14}\left(\frac{R_S}{\lambda}\right)^2 \right], \\ \nonumber
& & 8\pi\theta \left( -2q_1+\frac{a_3q_1}{a_2^2} - \frac{4}{3}\frac{q_2}{a_2} \right) \Sigma_5 =
\frac{e^{-r/\lambda}}{r}\frac{(GM_S)^2}{R_S}
\times \\ \nonumber && \theta \left[ \theta (\mu-2) -\frac{4}{3}\nu \right]
\left[ \left(\frac{\lambda}{R_S}\right)^2 + \frac{1}{10} + \frac{1}{280}\left(\frac{R_S}{\lambda}\right)^2 \right].
\end{eqnarray}
The following potentials contain both a Yukawa term and other functions of $r$:
\begin{widetext}
\begin{eqnarray}
&& \frac{1}{18\pi} \frac{(1-\theta)}{c^4} \frac{1}{a_2} X(U\Y) = \\ \nonumber &&
\frac{2}{3c^2} GM_S \frac{e^{-r/\lambda}}{r} (1-\theta) \frac{GM_S}{c^2R_S}
\left[ \left(\frac{R_S}{\lambda}\right)\mbox{Ei}\left(-2\frac{R_S}{\lambda}\right) +
\frac{34}{35}\left(\frac{R_S}{\lambda}\right)^2\right] + \\ \nonumber
&&\frac{2}{3c^2} \frac{GM_S}{r}\left[ e^{-r/\lambda}\ln\left(\frac{r}{R_S}\right) -
e^{r/\lambda}\mbox{Ei}\left(-2\frac{r}{\lambda}\right) \right](1-\theta)\frac{GM_S}{c^2R_S}
\left(\frac{R_S}{\lambda}\right), \\ && \frac{1}{36\pi} \frac{(1-\theta)^2}{c^4}\left(1+\frac{a_3}{2a_2^2}\right)\frac{1}{a_2} X(\Y^2) = \\ \nonumber
&& \frac{2}{3c^2} GM_S \frac{e^{-r/\lambda}}{r} (1-\theta)^2\left(1+\frac{\mu}{2}\right)\frac{GM_S}{c^2R_S} \left\{ \frac{1}{2}\left(\frac{R_S}{\lambda}\right)\left[\mbox{Ei}\left(-3\frac{R_S}{\lambda}\right)-
\mbox{Ei}\left(-\frac{R_S}{\lambda}\right)\right] + \frac{17}{35}\left(\frac{R_S}{\lambda}\right)^2 \right\} + \\ \nonumber
&& \frac{1}{2c^2} \frac{GM_S}{r}\left[ e^{-r/\lambda}\mbox{Ei}\left(-\frac{r}{\lambda}\right) -
e^{r/\lambda}\mbox{Ei}\left(-3\frac{r}{\lambda}\right) \right] (1-\theta)^2\left(1+\frac{\mu}{2}\right)\frac{GM_S}{c^2R_S}\left(\frac{R_S}{\lambda}\right),
\end{eqnarray}
\end{widetext}
where $\mbox{Ei}(x)$ denotes the exponential integral function, Eq. (\ref{Eifunction}).

The potential $\V(\Y^2)$ is decomposed into the sum of two
potentials $\V_1(\Y^2)+\V_2(\Y^2)$ where $\V_1$ is proportional to $1\slash r$ for $r>R_S$
(it is absorbed into the effective mass term), while $\V_2$ contains the following functions of $r$:
\begin{eqnarray}
& & \frac{1}{27\pi} \frac{(1-\theta)^2}{c^4} \frac{1}{a_2} \V_2(\Y^2) = \\ \nonumber
& & \frac{4}{9c^2} GM_S (1-\theta)^2 \frac{GM_S}{c^2R_S}\left(\frac{R_S}{\lambda}\right) \left[ \frac{e^{-2r/\lambda}}{r} + \frac{2}{\lambda} \mbox{Ei}\left(-2\frac{r}{\lambda}\right) \right].
\end{eqnarray}
For a static, spherically symmetric mass density $\rho=\rho(r)$ we find that, for $r>R_S$,
\begin{equation}
\psi_1(r) = \psi_2(r) = \psi_3(r) = 0 .
\end{equation}
Using the results in Appendix B, the linear combination of potentials $\psi_4$ and $\psi_5$ in $g_{00}$
is proportional to a Yukawa integral of the type $X(Q)$, with $Q$ supported inside the spherical body, so that
such a linear combination is proportional to a Yukawa term:
\begin{eqnarray}
& & -\frac{4}{3} \frac{(1-\theta)^2}{c^4} a_2 \left( \sqrt{\frac{2}{3a_2}} \psi_4 +2\psi_5 \right) = \\ \nonumber
& & -\frac{4}{45c^2} GM_S \frac{e^{-r/\lambda}}{r} (1-\theta)^2 \frac{GM_S}{c^2R_S} \left[ 1 + \frac{1}{14}\left(\frac{R_S}{\lambda}\right)^2 \right].
\end{eqnarray}
Analogously, the linear combination of potentials $\psi_6,\dots,\psi_9$ is also proportional to a Yukawa term:
\begin{eqnarray}
\nonumber & & \frac{4}{3} \frac{(1-\theta)^2}{c^4} a_2 \left[ 2\psi_6 +\sqrt{\frac{2}{3a_2}}\left(\psi_7+\psi_8\right) +
\frac{1}{3a_2} \psi_9 \right] = \\ \nonumber
& & \frac{2}{15c^2} GM_S \frac{e^{-r/\lambda}}{r} (1-\theta)^2 \frac{GM_S}{c^2R_S} \\
& & \times\left[ 1 -\frac{5}{9}\left(\frac{R_S}{\lambda}\right) + \frac{43}{210}\left(\frac{R_S}{\lambda}\right)^2 \right].
\end{eqnarray}
Eventually for the potential $\psi_{10}$ we find:
\begin{eqnarray}
\nonumber && \frac{2}{3a_2} \frac{(1-\theta)}{c^4}\left( -2a_2 + q_1 +\frac{a_3q_1}{a_2^2} - \frac{2}{3} \frac{q_2}{a_2} \right)\psi_{10} = \\ \nonumber &&
\frac{4}{5c^2} GM_S \frac{e^{-r/\lambda}}{r} (1-\theta) \frac{GM_S}{c^2R_S}
\left( -2 + \theta +\theta\mu - \frac{2}{3} \nu \right) \times \\ && \left[ 1 -\frac{5}{6}\left(\frac{R_S}{\lambda}\right) + \frac{11}{21}\left(\frac{R_S}{\lambda}\right)^2 \right].
\end{eqnarray}

\section*{Acknowledgments}

The work of R.M. is partially supported by INFN (Istituto Nazionale di Fisica Nucleare, Italy), as part of the MoonLIGHT-2 experiment in the framework of the research activities of the Commissione Scientifica Nazionale n. 2 (CSN2).



\end{document}